\documentclass[%
 reprint,
 pra,
 floatfix,
 amsmath,amssymb,
 aps, widetext, superscriptaddress
]{revtex4-2}

\usepackage{xcolor}
\usepackage{hyperref}
\usepackage{graphicx}
\usepackage{dcolumn}
\usepackage{bm}
\usepackage{physics}
\usepackage{xspace}
\usepackage{chemformula}
\usepackage{csquotes}
\usepackage{todonotes}
\newcommand{\opsigma}{\hat{\sigma}}
\newcommand{\Fe}{\ensuremath{{}^{57}\mathrm{Fe}}\xspace}

\newcommand{\SR}{\ensuremath{\Gamma_\mathrm{SR}}\xspace}
\newcommand{\CLS}{\ensuremath{\Delta_\mathrm{CLS}}\xspace}

\newcommand{\mb}{M\"ossbauer\xspace}

\newcommand{\tgamma}{\ensuremath{\tilde{\gamma}}}

\renewcommand{\vec}[1]{\ensuremath{{\bm{#1}}}}

\allowdisplaybreaks[3]
\begin{document}
\title{Inverse design in nuclear quantum optics: From artificial x-ray multi-level schemes to spectral observables}

\author{Oliver Diekmann}
\email{oliver.diekmann@mpi-hd.mpg.de}

\affiliation{%
 Max-Planck-Institut f\"ur Kernphysik, Saupfercheckweg 1, 69117 Heidelberg, Germany
}%
\affiliation{%
Institute for Theoretical Physics, Vienna University of Technology (TU Wien), Vienna A-1040, Austria
}

\author{Dominik Lentrodt}

\affiliation{%
 Max-Planck-Institut f\"ur Kernphysik, Saupfercheckweg 1, 69117 Heidelberg, Germany
}
\affiliation{Physikalisches Institut, Albert-Ludwigs-Universit\"at Freiburg, Hermann-Herder-Stra{\ss}e 3, D-79104 Freiburg, Germany}
\affiliation{EUCOR Centre for Quantum Science and Quantum Computing, Albert-Ludwigs-Universit\"at Freiburg, Hermann-Herder-Stra{\ss}e 3, D-79104 Freiburg, Germany}

\author{J\"org Evers}
\email{joerg.evers@mpi-hd.mpg.de}

\affiliation{%
 Max-Planck-Institut f\"ur Kernphysik, Saupfercheckweg 1, 69117 Heidelberg, Germany
}%

\date{\today}

\begin{abstract}
Ensembles of M\"ossbauer nuclei embedded in thin-film cavities form a promising platform for x-ray quantum optics. A key feature is that the joint nuclei-cavity system can be considered as an artificial x-ray multi-level scheme in the low-excitation regime. Using the cavity environment, the structure and parameters of such level schemes can be tailored  beyond those offered by the bare nuclei. However, so far, the direct determination of a cavity structure providing a desired quantum optical functionality has remained an open challenge. Here, we address this challenge using an inverse design methodology. As a first qualitative result, we show that the established fitting approach based on scattering observables in general is not unique, since the analysis may lead to different multi-level systems for the same cavity if based on observables in different scattering channels. Motivated by this, we distinguish between scattering signatures and the microscopic level scheme as separate design objectives, with the latter being uniquely determined by an \textit{ab initio} approach. We find that both design objectives are of practical relevance and that they complement each other regarding potential applications. We demonstrate the inverse design for both objectives using example tasks, such as realising electromagnetically induced transparency. Our results pave the way for new applications in nuclear quantum optics involving more complex x-ray cavity designs.
\end{abstract}

\maketitle

\section{Introduction}\label{sec:Introduction}

Combining many individual quantum systems to form a collective one with enhanced properties that yet features a simple description is a ubiquitous phenomenon, with prominent examples including the archetypal Dicke model~\cite{dicke_coherence_1954,GROSS1982301,doi:10.1080/09500340.2016.1215564,PRXQuantum.3.010201} or ``superatoms'' facilitated, e.g., by the Rydberg blockade~\cite{gallagher_rydberg_1994,lukin_dipole_2001}.
In nuclear x-ray quantum optics, this approach is especially attractive, as it allows one to implement versatile multi-level schemes which otherwise may be inaccessible in individual nuclei. In particular, the crucial requirement of external control fields, which are typically not available in present-day experiments with nuclei, can be alleviated by suitably engineering intrinsic collective couplings.

\begin{figure*}
\includegraphics[scale=0.95]{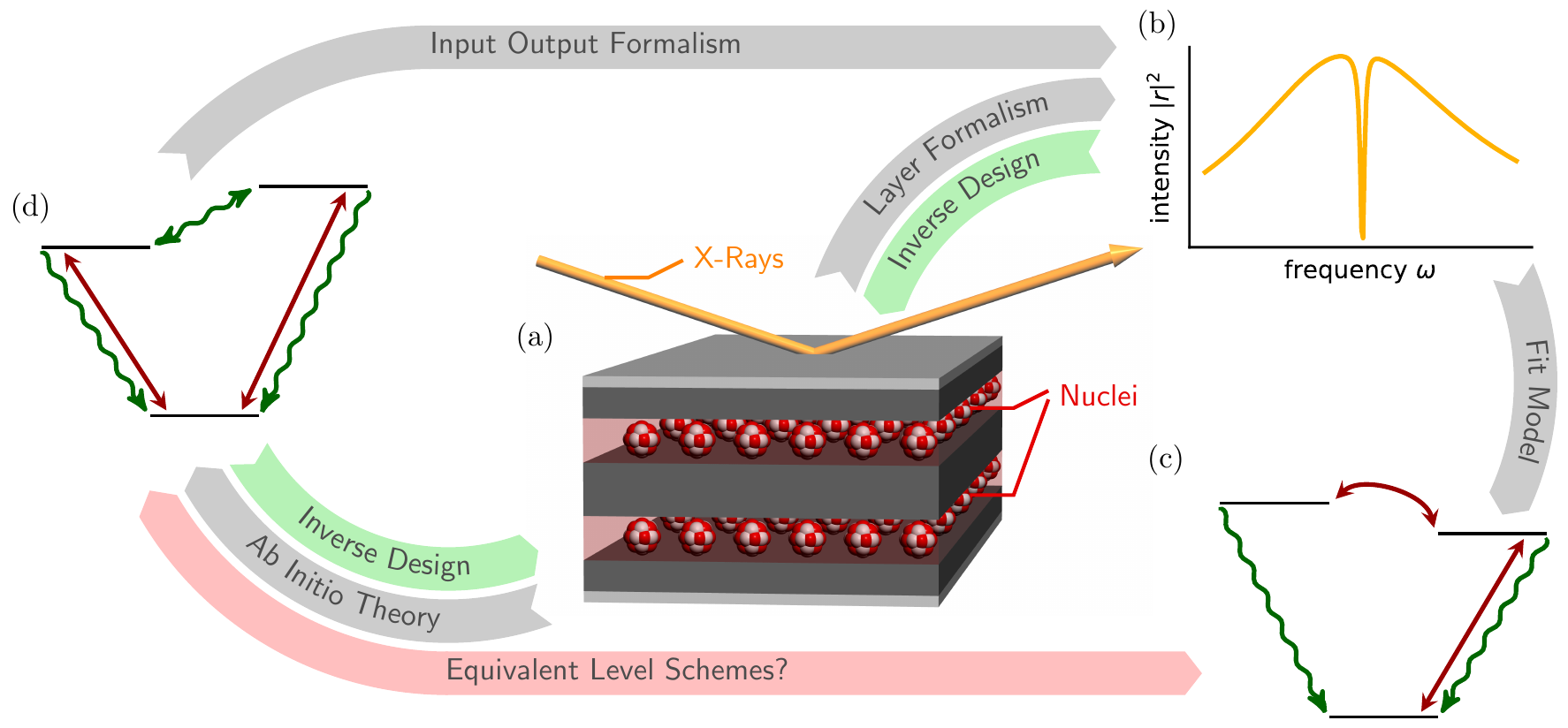}
\caption{Interpretation of thin-film cavities containing thin layers of  \mb nuclei in terms of multi-level scheme spectra. (a) Schematic example layout of a thin-film cavity with two resonant layers containing \mb nuclei. (b) The cavity can be probed using near-resonant x-rays, giving rise, e.g., to observable reflection spectra. In the figure, a reflection spectrum featuring a narrow resonance reminiscent of electromagnetically induced transparency is shown as an example. In the low-excitation limit, the cavity with nuclei can be interpreted in terms of an equivalent artificial multi-level quantum system~\cite{rohlsberger_electromagnetically_2012,heeg_x-ray_2013, heeg_collective_2015, lentrodt_ab_2020,yoshida_quantum_2021}. (c) This level scheme may be determined from the observable spectrum, e.g., by comparing or fitting the spectrum to the optical response of a multi-level model system~\cite{rohlsberger_electromagnetically_2012,heeg_x-ray_2013}. (d) Alternatively, a multi-level scheme can also be determined directly from the cavity structure, using an \textit{ab initio} theory~\cite{lentrodt_ab_2020}. Here, we show that the multi-level systems obtained via the two approaches in general are not equivalent [schematically illustrated by different level schemes in (c) and (d)], and that they complement each other in the applications of the cavity structure. Motivated by this result, we develop two different inverse design approaches~\cite{bennett_inverse_2020,diekmann_inverse_2022} which allow one to reverse the established methods of calculating the spectrum and the multi-level system from the cavity structure. Using these design tools, one can determine cavity structures which either satisfy particular constraints on the observable spectrum, or the artificial multi-level system, in an optimal way. This paves the way for new applications involving more complex x-ray cavity designs.
}
\label{fig:interrelations}
\end{figure*}

A successful x-ray platform for such artificial quantum systems comprises layers of \mb nuclei embedded in thin-film cavities (see Fig.~\ref{fig:interrelations}(a) and \cite{yoshida_quantum_2021} for a review). The stratified media~\cite{hannon_impedance-matched_1979, baron_angular_1994,toellner_observation_1995, andreeva_time-differential_1996,rohlsberger_theory_1999} can be considered cavity QED systems and --- when weakly probed in grazing incidence --- feature a description in terms of artificial level schemes. As a result, they have recently raised considerable interest as a platform for nuclear x-ray quantum optics~\cite{adams_nonlinear_2012,kuznetsova_quantum_2017,adams_scientific_2019,yoshida_quantum_2021}, both experimentally~\cite{rohlsberger_collective_2010,rohlsberger_electromagnetically_2012,heeg_vacuum-assisted_2013,haber_collective_2016,haber_rabi_2017,heeg_tunable_2015} and theoretically~\cite{heeg_x-ray_2013,heeg_collective_2015,kong_greens-function_2020,lentrodt_ab_2020,diekmann_inverse_2022,andrejic_superradiance_2021,huang_spacing-dependent_2020,fujiwara_quantum_2021}. Note that similar x-ray cavities have also been considered in the context of electronic resonances~\cite{haber_spectral_2019,huang_controlling_2021,vassholz_observation_2021, huang_classical_2022}.

The quantum systems comprising the building blocks of this platform are M\"ossbauer nuclei, which feature well-defined properties. However, in the low-excitation regime, only few collective excitations in the large interacting ensemble constitute the relevant quantum states. The set of these states can then be considered as an artificial level scheme. The collective nature of the description in the low-excitation regime becomes manifest in parameters such as level shifts \cite{rohlsberger_collective_2010,shv,friedberg_frequency_1973}, enhanced decay rates \cite{hannon_coherent_1999} and coupling constants \cite{rohlsberger_electromagnetically_2012,heeg_vacuum-assisted_2013,heeg_tunable_2015,haber_rabi_2017,heeg_collective_2015}, which are proportionally enhanced by the number density of participating nuclei and tunable via the cavity environment. Even for bare two-level nuclei, the effective multi-level description may comprise multiple excited states and coherent couplings between these states, which can compensate the hitherto lack of suitable multi-color x-ray control fields. Consequently, advanced quantum optical setups at x-ray energies have become accessible within the platform~\cite{yoshida_quantum_2021}.

This progress raises the question, which quantum optical schemes can in general be implemented using such thin-film cavity setups and which functionalities they can provide. To answer this question, it is necessary to first understand how the artificial multi-level scheme can be determined for a given cavity structure. So far, mainly two approaches have been used to characterize the level scheme. The first approach is based on fitting model predictions to the linear reflection spectrum [see Figs.~\ref{fig:interrelations}(b) and \ref{fig:interrelations}(c)], which is a primary observable in experiments and calculable using semiclassical scattering theory  (see Fig. 1 in \cite{lentrodt_ab_2020} for an overview of different methods). One may then compare the cavity reflectance to related optical properties of atomic systems, such as the susceptibility, in order to assess the realized level scheme~\cite{rohlsberger_collective_2010,rohlsberger_electromagnetically_2012,heeg_vacuum-assisted_2013,haber_collective_2016,heeg_x-ray_2013,heeg_collective_2015,kong_greens-function_2020,lentrodt_ab_2020}.
A second approach is to characterize the system by the relevant collective excitations of the nuclei coupled by the cavity -- the multi-level scheme -- and determine their properties [see Fig.~\ref{fig:interrelations}(d)]. The latter can be calculated on a microscopic~\cite{heeg_x-ray_2013, heeg_collective_2015, kong_greens-function_2020} or even \textit{ab initio}~\cite{lentrodt_ab_2020} basis using quantum optical techniques. Observables such as the linear reflection spectrum can then be derived in a second step from the multi-level system's parameters using input-output relations.

It has been shown that if one only considers the linear reflection spectrum, then the two approaches are in full agreement~\cite{lentrodt_ab_2020}.  
So far, however, it has not been explored whether the {\it interpretation} of the involved physical mechanisms on the basis of the linear reflection spectrum and microscopic considerations lead to the same conclusions on the underlying level scheme.  As one key result of the present work, we show that this in general is not the case. A concrete problem in this regard already arises from the fact that usually a fit model has to be chosen in order to determine the level scheme from the linear reflection spectrum, see Fig.~\ref{fig:interrelations}, and that there are different scattering observables. Therefore, already the observable linear spectra do not necessarily lead to a uniquely defined level scheme. Furthermore, even though the \textit{ab initio} approach leads to a unique level scheme~\cite{lentrodt_ab_2020}, it in general does not coincide with those obtained from observable spectra.

On the one hand, this finding is  of interest from a fundamental point of view, e.g., when the realization of a quantum optical effect is to be certified on the basis of a level scheme. It also affects the question in how far the artificial level schemes realized in x-ray cavities relate to the corresponding ``true'' level schemes, e.g., in single atoms---in particular, upon variation of external parameters. On the other hand, the ambiguity in the interpretation becomes particularly relevant from a practical perspective for the design and optimization of artificial level schemes with three or more coupled states towards a particular functionality, such as the enhancement of nonlinear light-matter interactions~\cite{fleischhauer_electromagnetically_2005}. To this end, recently, we showed for the case of a single resonant layer in a thin-film cavity that the engineering of the resulting two-level scheme can be addressed by an inverse design approach~\cite{diekmann_inverse_2022,bennett_inverse_2020}. Given design goals such as particular parameters for the resulting two-level scheme, this approach allows one to directly determine cavity structures which satisfy these design goals. A successful inverse design of more complex multi-level systems therefore also requires the definition of suitable design goals. However, it is not clear a priori how such goals formulated in terms of the observable spectrum and in terms of the effective quantum optical description map to each other or if they lead to different cavity realizations and functionalities.

Here, we address these questions and develop the inverse design of x-ray cavities with multiple resonant layers. Such cavity structures allow for the implementation of advanced artificial multi-level systems with three or more states. As compared to the two-level case studied previously, qualitatively different phenomena and applications are expected due to the cavity-mediated couplings between the different excited states, which enter as new parameters into the problem. As a starting point for our work, we analyze a particular highly symmetrical cavity structure. As a key insight of this paper, we use this cavity to show that an interpretation of the realized multi-level system based on the reflection spectrum can be inconsistent with an analogous interpretation in terms of the transmission spectrum, while the interpretation obtained from the microscopic \textit{ab initio} theory is unique. Using a detailed analysis, we reveal the origin of these deviating interpretations, show that the \textit{ab initio} approach can consistently account for both observables, and thereby reconcile the different approaches. As a result, we conclude that the design of observable spectra and the design of the underlying microscopic multi-level system in general relate to different optimization goals. Regarding practical applications of the thin-film systems, however, both design approaches are of relevance and complement each other. Therefore, we approach the inverse design of the thin-film structures along two different routes.

First, we develop the inverse design of the artificial multi-level scheme realized in the thin-film cavity, i.e., the direct design of the multi-level system parameters using the \textit{ab initio} approach~\cite{lentrodt_ab_2020} [see Fig.~\ref{fig:interrelations}(d)]. To illustrate the method, we focus on the simplest case of three-level systems realized by cavities with two resonant layers, and provide a comprehensive overview of the accessible three-level system parameters, i.e., the various transition frequencies, cavity-induced coupling- and (cross-)decay rates, as well as the effective driving fields. The inverse design approach in particular allows us to optimize ratios between relevant level scheme parameters such as Rabi frequencies and decay rates, which govern the quantum optical functionality. This is a prerequisite to the realization of advanced quantum optical schemes at x-ray energies, which we exemplify by the design of nuclear electromagnetically induced transparency (EIT)~\cite{rohlsberger_electromagnetically_2012, fleischhauer_electromagnetically_2005,heeg_collective_2015,lentrodt_ab_2021,kong_greens-function_2020}.

Second, we discuss the direct inverse design of the linear reflection spectrum, i.e., without referring to the parameters of the underlying artificial multi-level scheme [see Fig.~\ref{fig:interrelations}(b)]. In contrast to the previous part, here, the experimental observables are directly optimized, without referring to an underlying level scheme for the design. This approach is motivated by the observation that for some applications, the observable spectrum is the crucial quantity, and not the underlying multi-level scheme. Examples include frequency filters~\cite{rohlsberger_grazing_1994}, or the manipulation of the x-ray group velocity via steep linear dispersions~\cite{heeg_tunable_2015}. For the case of artificial two-level schemes, the observational aspects could largely be accounted for by introducing a visibility criterion~\cite{diekmann_inverse_2022}. However, with multi-level systems, the design of the observable spectra separates from that of the level scheme, for example, since different features of the level scheme may express themselves to a different extent in the overall reflection spectrum. To simplify the design, we rewrite the spectral response of the system in a diagonal basis~\cite{hannon_coherent_1999}, thereby disentangling the effect of the inter-layer couplings on the visible signatures. Using this approach, we demonstrate the possibility of tailoring the reflection spectrum towards specific requirements, e.g., by tuning the splitting of two spectral features, their absolute positioning, or their relative amplitude in the spectrum. We further show that the spectrum of a two-layer cavity with three relevant states can be designed to resemble the one of a two-level system, however, with a tunability of the two-level system parameters exceeding that of the corresponding single-layer case.

The work is structured as follows. In Sec.~\ref{sec:micromacro} we use an instructive example to illustrate differences in the interpretation of the thin-film structures. In Sec.~\ref{sec:fewLevel} we revise the multi-level description on the \textit{ab initio} basis and the calculation of the observable spectra, and introduce the spectrum's decomposition into a diagonal basis. In Sec.~\ref{sec:analysisSymmetric} we in-detail explain the observations of Sec.~\ref{sec:micromacro} using this \textit{ab initio} approach. In Sec.~\ref{sec:invth} we demonstrate the inverse design of artificial three-level systems realized by systems of two resonant layers within the \textit{ab initio} theory. In Sec.~\ref{sec:reflection} we exemplify the applicability of the inverse design to the reflection spectrum. Finally, we summarize our results and outline future research directions in Sec.~\ref{sec:conc}.

\section{Multi-level-system interpretation and inverse design objectives\label{sec:micromacro}}
\begin{figure*}[t]
%%\begin{captionbeside}[] %
\includegraphics[width=0.95\textwidth]{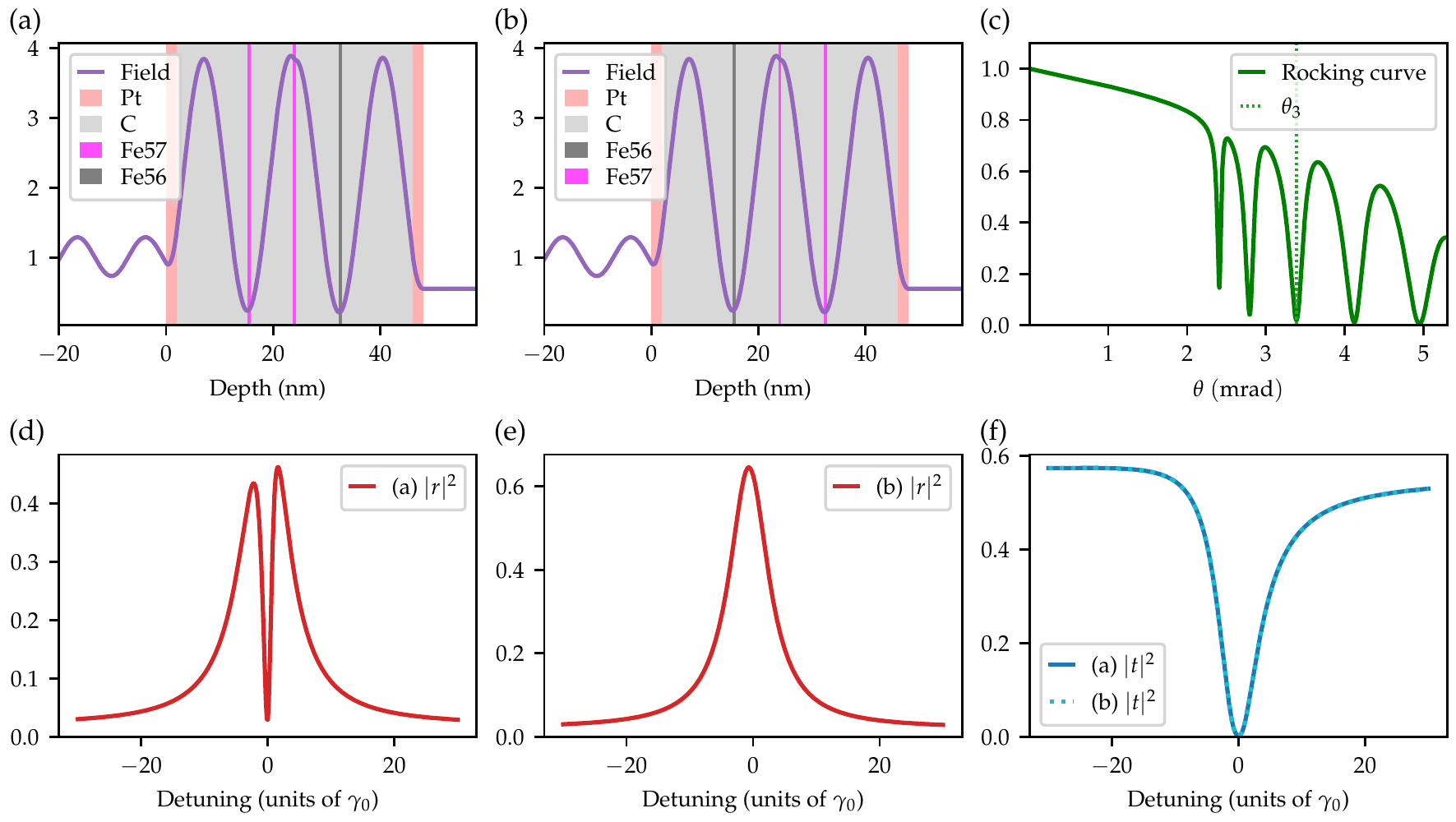}
\caption{\label{fig:eitNonEitDominik}Two model cavity setups used to motivate the two different inverse design approaches. Each cavity structure (a,b) comprises three iron layers symmetrically placed around the center of the cavity. Two of them are 
${}^{57}$Fe layers featuring the nuclear M\"ossbauer resonance, while the third one is a ${}^{56}$Fe layer without the nuclear resonance. In (a), the  
Pt($2.0~\mathrm{nm}$)/\allowbreak C($13~\mathrm{nm}$)/\allowbreak\Fe{}($0.57~\mathrm{nm}$)/\allowbreak C($8.0~\mathrm{nm}$)/\allowbreak\Fe{}($0.57~\mathrm{nm}$)/\allowbreak C($8.0~\mathrm{nm}$)/\allowbreak ${}^{56}\mathrm{Fe}${}($0.57~\mathrm{nm}$)/\allowbreak C($13~\mathrm{nm}$)/\allowbreak Pt($2.0~\mathrm{nm}$) structure has two resonant layers on the side where the probing x-rays impinge (left in the figure). The cavity in (b) is identical to the cavity in (a), but mirrored at the plane at half depth. The cavities therefore effectively only differ in the direction of incidence of the probing x-rays, but not in their off-resonant electronic properties. The two panels (a,b) further show the off-resonant x-ray intensity distributions normalized to the incident intensities, which agree with each other.
(c) shows the off-resonant electronic rocking curve ($\theta-2\theta$ scan) of the two cavities as a function of the incidence angle $\theta$, which also agree.
(d) and (e) show the nuclear reflection spectra of the two cavities in (a) and (b), respectively. They are evaluated at the incidence angle $\theta_3 = 3.39~\mathrm{mrad}$
 corresponding to the third minimum of the rocking curve indicated in (c), as for the intensity patterns in (a) and (b).
(f) shows the corresponding transmission spectrum of the cavities (a,b).
While the reflection spectra in (d) and (e) are different, the transmission spectra in (f) coincide, thus leading to potentially different phenomenological interpretations for the underlying level scheme.}
%It can be seen that the two reflection spectra (d,e) are different. Consequently, the interpretation of the cavities as artificial  multi-level schemes on the basis of the reflection spectra will lead to different level schemes. In contrast, the transmission spectra (f) agree, hence, leading to an interpretation in terms of the same level scheme. This illustrates that the interpretation of the cavity in terms of observable spectra is not unique for a given cavity structure. In contrast, the multi-level system derived from the ab initio approach only depends on the off-resonant electronic structure and therefore is unique, but in general may differ from the interpretations in terms of the observable spectra.}%
%\end{captionbeside}
\end{figure*}
 
In this Section, we illustrate the interpretation of a given cavity structure in terms of an artificial multi-level scheme. In particular, we show how the interpretation based on different observables and on the microscopic \textit{ab initio} theory~\cite{lentrodt_ab_2020} may lead to different conclusions regarding the artificial multi-level scheme formed by the nuclei-cavity system. This observation motivates the two different inverse design approaches developed in the following sections, based on the optimization of the multi-level scheme and on the optimization of the observable spectra, respectively.
As we are mostly interested in qualitative differences throughout this section, we restrict the discussion to a largely non-technical level. Detailed calculations substantiating the results of this part will then follow in Sec.~\ref{sec:analysisSymmetric}.

\subsection{Exemplary system}
For a qualitative analysis, we consider the two cavity structures shown in Fig.~\ref{fig:eitNonEitDominik}(a,b). Each of the two  cavities contains three iron layers, placed symmetrically around the plane at half depth. The iron layers comprise two different isotope species: The M\"ossbauer isotope ${}^{57}$Fe features a nuclear resonance at 14.4~keV transition energy. In contrast, the non-M\"ossbauer isotope ${}^{56}$Fe does not have a narrow nuclear resonance, but otherwise is equivalent to ${}^{57}$Fe in its electronic properties.
In Fig.~\ref{fig:eitNonEitDominik}(a), the two layers closer to the side at which the x-rays enter the cavity (towards negative depths) are made from ${}^{57}$Fe, while in the cavity in (b)  the two iron layers further away from the x-ray entry layer comprise ${}^{57}$Fe. Thus, the cavity in Fig.~\ref{fig:eitNonEitDominik}(a) is the mirror structure of the one in (b) with respect to the plane at half depth. 
Since the two cavities only differ in the isotope enrichment of the three iron layers, their electronic properties are the same. As a result, both are equivalent at off-resonant energies. Here, off-resonant refers to x-ray frequencies slightly detuned from the nuclear resonance in the M\"ossbauer isotope ${}^{57}$Fe, such that the electronic properties dominate the response. Furthermore, off-resonance, each of the two cavities itself is mirror-symmetric with respect to its respective plane at half depth. 
This highly symmetric cavity layout will allow us to qualitatively compare the different approaches to interpret the cavity functionality. Note that in order to preserve this symmetry, we do not consider a substrate on which such cavities are usually deposited during the fabrication. 

The choice of the structures in Fig.~\ref{fig:eitNonEitDominik}(a,b) is motivated by cavities with two resonant layers which may feature an EIT-like signature in the reflection spectrum, i.e., a broad background Lorentzian line shape with a narrow pronounced central dip [cf. Fig.~\ref{fig:eitNonEitDominik}(d)]~\cite{fleischhauer_electromagnetically_2005,rohlsberger_electromagnetically_2012, heeg_collective_2015, lentrodt_ab_2020}.
Based on the observable reflection spectrum, it was found in~\cite{rohlsberger_electromagnetically_2012} that suitable systems with two resonant layers can be interpreted in terms of a $\Lambda$-type three-level system known to exhibit EIT~\cite{fleischhauer_electromagnetically_2005}.
It was further found that EIT-like spectra only appear  if the off-resonant  electric field configuration inside the   cavity is such that it shows a node at the position of the resonant layer closer to the impinging x-rays and an antinode at the one further away [cf.  Fig.~\ref{fig:eitNonEitDominik}(a)].  In contrast, inverting the order towards an antinode-node configuration [cf. Fig.~\ref{fig:eitNonEitDominik}(b)], the EIT-like signature vanishes. The qualitatively different spectra were attributed to a vanishing control field between the two excited states in the antinode-node configuration in~\cite{rohlsberger_electromagnetically_2012}, and subsequently, much work has been devoted to the understanding of the appearance and non-appearance of EIT-like signatures for two-layer cavities~\cite{heeg_x-ray_2013, heeg_collective_2015, hu_electromagnetically-induced-transparencylike_2015, huang_field_2017, lentrodt_ab_2020,kong_greens-function_2020}.

For our discussion, we investigate the difference between EIT and non-EIT signatures for the above-introduced cavities. Their symmetry implies exact relations between the two realized level schemes, which provide insight into the underlying mechanisms. As expected from this symmetry, the off-resonant intensity distributions of the electric fields in (a,b) agree. The cavities also share the same off-resonant electronic reflection intensity as a function of the incidence angle $\theta$, as shown in panel (c). In this \textit{rocking curve}, the dips in the reflection intensity correspond to the different cavity modes. The off-resonant field intensity distributions in (a,b) were calculated for an incidence angle $\theta_3$ corresponding to the third cavity mode.

\subsection{Multi-level-system interpretation of scattering observables}
We are now in the position to interpret the two cavities in terms of their resonant nuclear reflection spectra, shown in Fig.~\ref{fig:eitNonEitDominik}(d) and (e), respectively. We find that the cavity in (a) with spectrum in (d) exhibits an EIT-like feature, while cavity (b) with spectrum (e) does not. These results are consistent with those in~\cite{rohlsberger_electromagnetically_2012}, since the resonant layers in (a) are placed in a node-antinode (from left to right) configuration, while (b) realizes an antinode-node configuration. The observable reflection spectrum thus suggests that the cavity in (a) corresponds to a three-level system exhibiting EIT, while the cavity (b) has a different multi-level representation that does not feature EIT.

Next, we compare this result with the analogous analysis of the cavity transmission spectra. As shown in (f), the resonant nuclear transmission spectra of the two cavities (a,b) coincide, and do not feature narrow resonances reminiscent of EIT.  This agreement of the transmission spectra suggests an interpretation of the collective nuclear behavior in terms of equivalent level schemes for the cavities in Fig.~\ref{fig:eitNonEitDominik}(a,b). From this point of view, the multi-level interpretation of the cavity based on the transmission spectra appears to be at odds with the corresponding one obtained from the reflection spectra.

\subsection{Multi-level-system interpretation of the nuclear dynamics in the cavity}
As a complementary approach, we can analyze the artificial level scheme on the basis of the \textit{ab initio} theory~\cite{lentrodt_ab_2020}. It does not rely on the characterization via observables, but instead derives the  artificial level scheme and the master equation governing the nuclear dynamics directly from the cavity geometry. We will provide technical details on this approach in Sec.~\ref{sec:fewLevel}, and a detailed analysis of the situation in Fig.~\ref{fig:eitNonEitDominik} in Sec.~\ref{sec:analysisSymmetric}.
Here, it suffices to consider that the cavity-induced coupling constants, decay rates and level shifts entering the nuclear master equation depend only on the off-resonant electronic characteristics of the cavity. Since the off-resonant electronic structures of the two cavities in (a,b) are the same, one finds that the \textit{ab initio} descriptions for the two settings of Figs.~\ref{fig:eitNonEitDominik}(a) and \ref{fig:eitNonEitDominik}(d) are identical due to the inherent symmetry between both settings. Thus, the level schemes are equivalent. However, in the following more detailed analysis in Secs.~\ref{sec:theo} and~\ref{sec:analysisSymmetric} we will find that the different placements of the resonant layers also affects the incoupling of the external x-ray field, such that the nuclei in cavity (a) experience different driving fields (Rabi frequencies) from those in cavity (b).  These different effective driving fields experienced by the nuclei account for the different reflection spectra of cavities (a,b), even though the decay rates, level shifts and inter-level couplings rates of the level schemes in (a,b) are identical. We note that this latter insight is different from the conclusions based on the observable reflection spectrum, which instead suggests a vanishing coupling between the two excited states in the multi-level description of the cavity (b) as the origin of the different spectra. 

Within the \textit{ab initio} description also the transmission spectra can be analyzed. They are the same for cavities (a,b), even though the Rabi frequencies of the multi-level schemes for (a,b) differ, and thus the reflection spectra. This can be understood by noting that also the outcoupling of the field generated by the nuclei into the observation channel needs to be considered. For cavities (a,b), the outcouplings deviate in such a way that they cancel the differences in the incoupling, as will be shown quantitatively in Sec.~\ref{sec:analysisSymmetric}. Summarizing, the observable reflection and transmission spectra are a result of the \textit{ab initio} multi-level scheme in combination with the in- and outcoupling contributions. If the electromagnetic environment is the same for two settings, then differences in the observable signatures can only stem from in- and outcoupling modifications. Within the \textit{ab initio} theory~\cite{lentrodt_ab_2020} it can therefore be explained consistently that the transmission spectra of (a,b) coincide while the reflection spectra do not.

\subsection{Two complementary inverse design objectives}
The analysis of the example cavity in Fig.~\ref{fig:eitNonEitDominik} shows that the observable spectra and the \textit{ab initio} approach may give rise to different interpretations in terms of an underlying multi-level scheme. This distinction is of relevance for defining criteria for the certification of the observation of particular quantum optical effects, which may either be based on the properties of the observable spectra directly, or on the nuclear dynamics via the underlying {\it ab initio} level scheme.

Regarding the design of novel or improved functionalities in thin-film structures, we therefore conclude that both, the design of the microscopic level scheme following from the \textit{ab initio} approach and the direct design of observable spectra, are useful. In general, they complement each other while leading  to different optimum cavity settings. If a particular spectral or temporal response of the cavity is desired, e.g., for the design of filters~\cite{rohlsberger_grazing_1994}, to control the group velocity of x-ray pulses~\cite{heeg_tunable_2015}, or in general  to shape an x-ray pulse in a particular way, then an optimization based on the observable reflection spectrum may be favorable. In this case, the particular multi-level scheme realized by the cavity may not even be of relevance. 
On the other hand, when aiming for applications that rely on a particular level scheme, e.g., to transfer quantum optical effects to the x-ray regime, an optimization of the multi-level scheme in terms of the \textit{ab initio} theory~[25] appears favorable---in particular, if experimental parameters are to be varied, such as the externally applied x-ray intensity. Then, level schemes associated to the observable spectra may not follow the usual quantum optical intuition, since they are based on a combination of the nuclear dynamics with the in- and outcoupling, which react to parameter changes in different ways. In contrast, the   \textit{ab initio}  multi-level scheme focuses on the nuclear dynamics only, and therefore in general has a more transparent parameter dependence.
%On the other hand, if one aims at realizing a particular level scheme, e.g., to implement a quantum optical effect,  then an optimization in terms of the multi-level parameters appears favorable. 
In this case, however, it is important to note that the observable signatures may not agree with those expected from the quantum optical treatment of atoms in free space, because of the influence of the outcoupling of the x-rays from the cavity.

Because of this complementary nature of the two design approaches, in the following, we will discuss  both inverse design cases separately.

\section{Theoretical model}\label{sec:theo}
In this Section we briefly revise the theoretical foundations of the description of \mb nuclei in thin-film cavities on the basis of the \textit{ab initio} theory~\cite{lentrodt_ab_2020,asenjo-garcia_atom-light_2017} and the previous inverse design approach~\cite{diekmann_inverse_2022}. We start out considering the multi-level scheme description. Subsequently, we revisit the calculation of the experimentally observable spectra and introduce a diagonal basis to disentangle the different contributions to these observables.
 
\subsection{\textit{Ab initio} multi-level description\label{sec:fewLevel}}
Our discussion is based on the treatment of layers of \mb nuclei in thin-film cavities as quantum multi-level schemes in the regime of low excitation. 
For simplicity, we assume the absence of hyperfine splittings in the individual \mb nuclei. Thus, they are well approximated as two-level systems with bare transition frequency $\omega_\mathrm{nuc}$. 
Following the derivation in~\cite{lentrodt_ab_2020}, the master equation governing the dynamics of the artificial multi-level quantum system is
\begin{equation}
    \dot{\rho} = -i\comm{\hat{H}}{\rho} + \mathcal{L}[\rho]\,.
    \label{eq:Master}
\end{equation}
The respective Hamiltonian can be written as
\begin{align}
    \hat{H}=\sum\limits_l\frac{\omega_{\mathrm{nuc}}}{2}\opsigma^z_{l}-\sum\limits_{ll'}\Delta_{ll'}{\opsigma}^+_{l}{\opsigma}^-_{l'}
    -\sum\limits_l\left(\Omega_l{\opsigma}^+_{l}+\mathrm{H.c.}\right)\,,\label{eq:single1}
\end{align}
where we employ natural units, $\hbar=c=1$. Here, the index $l$ labels the different excited states of the multi-level system. For the considered bare two-level nuclei, each excited state corresponds to a collective spin-wave excitation involving the coherent superposition of nuclear excitations in a particular layer, described by the Pauli operators $\opsigma^z_{l}$ and $\opsigma^\pm_{l}$. The creation of such a collective excitation is mediated by $\opsigma^+_{l}$ while $\opsigma^{-}_{l}$ annihilates an excitation.  Noting that in the grazing incidence regime the reflection- and transmission coefficients for \textit{s} and \textit{p} polarization become identical~\cite{als-nielsen_elements_2011}, and that in the absence of magnetic splitting the scattering at the nuclear resonances leaves the polarization unchanged, we focus on \textit{s} polarized light throughout the paper and omit the vectorial nature of electric fields and related quantities. 

The artificial multi-level scheme description comprises dipole couplings
\begin{equation}
\Delta_{ll'}=\frac{N}{A}\mu_0\omega^2_{\mathrm{nuc}}|d|^2\Re\left[{G}(z_l, z_{l'}, \vec{k}_\parallel, \omega_\mathrm{nuc})\right]\label{eq:DeltaEff}
\end{equation}
between the spin-wave excitations. These couplings are mediated by the in-plane Fourier transformed electromagnetic Green's function, which is  analytically known for stratified media~\cite{tomas_green_1995}. The position arguments of the Green's function  are evaluated at the depths $z_{l}$ and $z_{l'}$ of the resonant layers in the stack. Note that the diagonal energy shifts $\Delta_{ll}$ of the layers have been interpreted in terms of a collective Lamb shift (CLS)~\cite{rohlsberger_collective_2010}.
In Eq.~(\ref{eq:DeltaEff}), the in-plane wave vector is determined by the incidence angle of the x-rays and the Green's function is evaluated at the resonance frequency of the nuclei $\omega_\mathrm{nuc}$. The dipole moments of the bare nuclear transition are denoted by ${d}$ and the in-plane number density of nuclei in a resonant layer is written as $N/A$.

The collective excitations are driven by
\begin{equation}
\Omega_l = \frac{N}{A}{d}^*{E}_\mathrm{in}(z_l, \vec{k}_\parallel)\,.
\end{equation}
Here, ${E}_\mathrm{in}(z_l, \vec{k}_\parallel)$ is the classical electric field at the depth $z_l$ of the respective nuclear ensemble Fourier transformed along the plane of the layers. It is modified from its free-space value by the different electronic reflection channels inside the cavity, and can be evaluated, e.g., using a recursive approach known as the Parratt's formalism in the x-ray literature~\cite{parratt_surface_1954}. Explicit expressions in frequency space are given in~\cite{tomas_green_1995}.

Besides the coherent dynamics, the incoherent contributions are described by a Lindblad term, 
\begin{align}
    \mathcal{L}[\rho]&=\sum\limits_{ll'}\frac{\gamma_{ll'}+\delta_{ll'}\gamma_0}{2}\left[2{\opsigma}^-_{l'}\rho{\opsigma}^+_{l}-\acomm{{\opsigma}^+_{l}{\opsigma}^-_{l'}}{\rho}\right]\,, \label{eq:single2}
\end{align}
with the Kronecker delta $\delta_{ll'}$.
The Lindbladian not only comprises direct decay channels with rates $\gamma_{ll}$ but also  cross-decay terms $\gamma_{ll'}$ due to the cavity-mediated nuclear couplings. The explicit values of these incoherent contributions are mediated by the imaginary part of the cavity's Green's function,
\begin{align}
\gamma_{ll'}&=2\frac{N}{A}\mu_0\omega^2_{\mathrm{nuc}}|{d}|^2\Im\left[{G}(z_l, z_{l'}, \vec{k}_\parallel , \omega_\mathrm{nuc})\right]
    \,. \label{eq:GammaEff}
\end{align}
Note that the decay terms in Eq.~\eqref{eq:single2} also comprise a term involving the bare nuclear decay rate $\gamma_0$ which is approximately equal to the decay rate by internal conversion. The diagonal contributions $\gamma_{ll}$ effectively increase the direct decay rates of the excitated multi-level states, which is known as superradiance (SR)~\cite{dicke_coherence_1954,hannon_coherent_1999}.

From this \textit{ab initio} multi-level model, the observable spectra can be calculated~\cite{lentrodt_ab_2020}, as is summarized in the next Section~\ref{sec:normalTheory}. Note that alternatively, the spectra can also be calculated directly, e.g., with the semiclassical layer formalism~(see \cite{rohlsberger_nuclear_2005} for a review). 
However, the \textit{ab initio} approach is favorable for the inverse design of the reflection spectrum,  as it does not require fitting procedures to extract the properties of the spectrum for the comparison with the design objectives.

\begin{table*}[t!]
\begin{tabular}{ccccc}
\hline
\hline
Cavity&Level scheme parameters&Rabi frequencies&Reflection outcoupling&Transmission outcoupling\\
\hline
Fig.~\ref{fig:eitNonEitDominik}(a)
&
{\footnotesize$\left(\begin{matrix}
-0.040+0.22i&-0.86+0.012i\\
-0.86+0.012i&0.64+3.5i\\
\end{matrix}\right)$}
&{\footnotesize$\left(\begin{matrix}
-0.37+0.34i\\
-1.6-1.2i\\
\end{matrix}\right)$}
&{\footnotesize$\left(\begin{matrix}
0.26-0.23i\\
1.1+0.80i\\
\end{matrix}\right)$}&
{\footnotesize$\left(\begin{matrix}
-0.24+0.23i\\
1.1+0.80i\\
\end{matrix}\right)$}\\
Fig.~\ref{fig:eitNonEitDominik}(b)
&{\footnotesize$\left(\begin{matrix}
-0.040+0.22i&-0.86+0.012i\\
-0.86+0.012i&0.64+3.5i\\
\end{matrix}\right)$}
&{\footnotesize$\left(\begin{matrix}
0.35-0.33i\\
-1.6-1.2i\\
\end{matrix}\right)$}
&{\footnotesize$\left(\begin{matrix}
-0.24+0.23i\\
1.1+0.80i\\
\end{matrix}\right)$}&
{\footnotesize$\left(\begin{matrix}
0.26-0.23i\\
1.1+0.80i\\
\end{matrix}\right)$}
\\
\hline
\hline
\end{tabular}
\caption{Level scheme parameters, Rabi frequencies (incoupling contributions) and outcoupling contributions in reflection and transmission for the two symmetric cavities in Figs.~\ref{fig:eitNonEitDominik}(a) and \ref{fig:eitNonEitDominik}(b). The convention is  such that the second entry in the vector refers to the central resonant layer for both settings.
In the table, the level scheme parameters $(\Delta_{ll'}+i\gamma_{ll'}/2)$ are non-dimensionalized by scaling with the bare nuclear decay rate $\gamma_0$, the Rabi-frequencies $\tilde{\vec{\Omega}}$ are scaled by $id^*N/A$ to have the units of the incident electric field, and the reflection $\vec{G}^{\mathrm{r}}$ and transmission $\vec{G}^{\mathrm{t}}$ outcouplings are non-dimensionalized by scaling with $A\gamma_0/(id^*\mu_0\omega_{\mathrm{nuc}}^2N)$.
We find the level scheme parameters to be the same for the two cavities, since the level scheme depends on the off-resonant field distribution inside the cavity.
The Rabi frequencies differ for the two cavities, because of the longer propagation of the x-rays to the deeper-lying resonant layers in cavity (b). 
Similarly, the reflection outcoupling and the transmission outcoupling depend on the required propagation pathway. While the absolute values for these parameters are similar for the two cavities, a characteristic sign flip distinguishes the first entries of the driving and outcoupling vectors of the two cavities.\label{tab:symmetric_nondiagonal}}
\end{table*}

\begin{table*}[t!]
\begin{tabular}{cccc}
\hline
\hline
Cavity&Eigenvalues&Reflection weights&Transmission weights\\
\hline
 Fig.~\ref{fig:eitNonEitDominik}(a)
&{\footnotesize$\left(\begin{matrix}
-0.090+0.45i\\
0.69+3.2i\\
\end{matrix}\right)$}
&{\footnotesize$\left(\begin{matrix}
-0.13+0.80i\\
-0.65-3.1i\\
\end{matrix}\right)$}&
{\footnotesize$\left(\begin{matrix}
-0.029+0.046i\\
-0.73-2.7i\\
\end{matrix}\right)$}\\
Fig.~\ref{fig:eitNonEitDominik}(b)
&{\footnotesize$\left(\begin{matrix}
-0.090+0.45i\\
0.69+3.2i\\
\end{matrix}\right)$}
&{\footnotesize$\left(\begin{matrix}
-0.0030+0.0021i\\
-0.77-2.4i\\
\end{matrix}\right)$}&
{\footnotesize$\left(\begin{matrix}
-0.029+0.046i\\
-0.73-2.7i\\
\end{matrix}\right)$}\\
\hline
\hline 
\end{tabular}
\caption{Eigenvalues $(\lambda_l)/\gamma_0$ of the level scheme parameter matrix $(\Delta_{ll'}+i\gamma_{ll'})$ and the respective weights in the reflection $(g^\mathrm{r}_l)/\gamma_0$ and transmission $(g^\mathrm{t}_l)/\gamma_0$ spectrum. The two cavities differ in the reflection spectrum, as in (a) both eigenmodes contribute strongly, whereas the contribution of the first eigenmode is suppressed in cavity (b). In contrast, the two transmission spectra agree, and share the same eigenmode weights.\label{tab:symmetric_diagonal}}
\end{table*}

\subsection{Spectral observables in the \textit{ab initio} approach}\label{sec:normalTheory}
From Eq.~\eqref{eq:Master} together with Eqs.~\eqref{eq:single1} and \eqref{eq:single2}, the equation of motion of $\sigma^-_l(t)\equiv\expval{\opsigma^-_{l}}$ can be obtained. In the low-excitation limit $\sigma^{z}_{l}(t)\equiv\expval{\opsigma^z_{l}}\approx -1$, it reads
\begin{align}
\dot{\sigma}^-_{l}(t) = &-i\left(\omega_\mathrm{nuc}-i\frac{\gamma_0}{2}\right)\sigma^-_{l}(t)\notag\\&+i\sum\limits_{l'}\left(\Delta_{ll'}+i\frac{\gamma_{ll'}}{2}\right)\sigma^-_{l'}(t)\notag\\&+ i\frac{N}{A}{d}^*{E}_{\mathrm{in}}(z_l, \vec{k}_\parallel)\,.
\end{align}
Using the Fourier transform, $\sigma^-_{l}(t)=\int\dd{\omega}e^{-i\omega t}\sigma^-_{l}(\omega)$, the dynamics is solved in frequency space as
\begin{align}
\vec{\sigma} = -\vec{\mathcal{M}}^{-1}\vec{\Omega}\,,
\end{align}
where 
\begin{align}
\left(\vec{\sigma}\right)_l &= \sigma^-_l(\omega)\,,\\
\left(\vec{\mathcal{M}}\right)_{ll'}&=i\left[\delta_{ll'}(\omega-\omega_\mathrm{nuc}+i\frac{\gamma_0}{2})+\Delta_{ll'}+i\frac{\gamma_{ll'}}{2}\right]\,,\label{eq:M}\\
\left(\tilde{\vec{\Omega}}\right)_l&=i\frac{N}{A}d^*E_\mathrm{in}(z_l, \vec{k}_\parallel, \omega_\mathrm{nuc})\,\label{eq:Om}.
\end{align}
Note  that for the calculation of the reflection spectrum, we approximate the frequency space electric field configuration $E_\mathrm{in}(z_l, \vec{k}_\parallel, \omega_\mathrm{nuc})$, that determines the Rabi frequencies $(\tilde{\vec{\Omega}})_l$, as constant over the narrow energy range of the  nuclear linewidth. Further, since all calculations are done in the linear regime, the electric field amplitude impinging on the cavity is normalized to unity throughout the paper. Consequently, $E_\mathrm{in}(z_l, \vec{k}_\parallel, \omega_\mathrm{nuc})$ describes the electric field enhancement at the respective nuclear layer.

Finally, by an input-output relation~\cite{asenjo-garcia_atom-light_2017, lentrodt_ab_2020}, the reflectance
\begin{align}
r(\omega) = r_\mathrm{el}-\mu_0\omega^2_{\mathrm{nuc}}\vec{G}^\mathrm{r}\vec{\mathcal{M}}^{-1}\tilde{\vec{\Omega}}\label{eq:reflection}
\end{align} 
can be calculated. Here, $r_\mathrm{el}$ describes the electronic (background) reflection, which stems from the x-ray light that is electronically reflected by the cavity structure. The nuclear contribution to the reflectance is given by the second term and contains the term $(\vec{G}^\mathrm{r})_l=d\,G(0, z_l, \vec{k}_\parallel, \omega_\mathrm{nuc})$, which propagates the nuclear response from the position of the nuclei to the cavity's upper surface. The squared absolute value of the reflectance $r(\omega)$ then denotes the experimentally accessible reflection spectrum. 

Analogously, the transmittance can be calculated,
\begin{align}
t(\omega) = t_\mathrm{el}-\mu_0\omega^2_{\mathrm{nuc}}\vec{G}^\mathrm{t}\vec{\mathcal{M}}^{-1}\tilde{\vec{\Omega}}\,.\label{eq:transmission}
\end{align}
Here, the electronic background transmission is denoted as $t_\mathrm{el}$ and the propagation of the nuclear response to the bottom surface of the cavity is mediated by $(\vec{G}^\mathrm{t})_l=d\,G(D, z_l, \vec{k}_\parallel, \omega_\mathrm{nuc})$ with $D$ the thickness of the cavity layer stack.

\subsubsection{Spectral eigenmodes}
By the matrix contractions in Eqs.~\eqref{eq:reflection} and \eqref{eq:transmission}, the spectra depend in a complex way on the properties of the different spin-wave excitations, their driving field as well as the outcoupling. Building on the above formalism, here, we introduce a separation of the nuclear response into its different contributions, using a diagonal basis. This will allow us to gain clearer access to features in the spectra. To that end, we assume the diagonalizability of the matrix $(\Delta_{ll'}+i\gamma_{ll'}/2)$. In this case, there exist matrices $\vec{S}, \vec{S}^{-1}$ such that $\vec{S}^{-1}(\Delta_{ll'}+i\gamma_{ll'}/2)\vec{S}$ is diagonal with complex eigenvalues $\lambda_l$. This transformation also diagonalizes $\vec{\mathcal{M}}^{-1}$ and we can hence write
\begin{align}
r(\omega) &= r_{\mathrm{el}}-\mu_0\omega_{\mathrm{nuc}}^2\left(\vec{{G^{\mathrm{r}}}}\vec{S}\right)\left(\vec{S}^{-1}\vec{\mathcal{M}}^{-1}\vec{S}\right)\left(\vec{S}^{-1}\tilde{\vec{{\Omega}}}\right)\notag\\ 
&=r_{\mathrm{el}}+\sum\limits_{l=1,2}\frac{ig^{\mathrm{r}}_l}{\omega-\omega_{\mathrm{nuc}}+i\frac{\gamma_0}{2}+\lambda_l}\,,\label{eq:diagonalizedSpectrum}\\
t(\omega) %&= t_{\mathrm{el}}-\mu_0\omega_{\mathrm{nuc}}^2\left(\vec{\tilde{G}}\vec{S}\right)\left(\vec{S}^{-1}\vec{\mathcal{M}}^{-1}\vec{S}\right)\left(\vec{S}^{-1}\vec{\tilde{\Omega}}\right)=\notag\\ 
&=t_{\mathrm{el}}+\sum\limits_{l=1,2}\frac{ig^{\mathrm{t}}_l}{\omega-\omega_{\mathrm{nuc}}+i\frac{\gamma_0}{2}+\lambda_l}\,.
\end{align}
The reflectance as well as the transmittance are given by a respective electronic background term, and by a Lorentzian contribution for each resonant layer. The width and position of each contribution are determined by the eigenvalue $\lambda_l$, and its weight is given by the product $g^{\mathrm{r (t)}}_l = \mu_0\omega_{\mathrm{nuc}}^2(\vec{G}^{\mathrm{r (t)}}\vec{S})_l(\vec{S}^{-1}\vec{\tilde{\Omega}})_l$ between the transformed driving and outcoupling vector components for the reflectance (transmittance). Note that these weights are frequency independent as the vectors as well as the transformations $\vec{S}$ do not depend on frequency. The frequency shift and linewidth enhancement of the Lorentzians are then directly accessible by the (negative) real and (positive) imaginary part of the eigenvalues, respectively.

All parameters appearing in the above equations can be calculated efficiently on the basis of the analytically available Green's functions and field configurations~\cite{tomas_green_1995}. The approach hence not only allows one to calculate the observables for a given cavity but also to efficiently implement the inverse design of the level scheme description as well as the spectra. 

In the following, we first apply the theoretical tools to understand the peculiarities in the multi-level interpretation of the cavity settings in Fig.~\ref{fig:eitNonEitDominik} found in Sec.~\ref{sec:micromacro}. Afterwards, we turn to the inverse design.

\section{\label{sec:analysisSymmetric}Origin of the differing multi-level-system interpretations}

With the artificial level scheme description and the diagonal representation of the spectra in Sec.~\ref{sec:theo}, we now have suitable tools to analyze the cavity setups presented in Sec.~\ref{sec:micromacro} quantitatively -- from the \textit{ab initio} level scheme perspective as well as regarding the resulting observable spectra.  

For both cavities, Figs.~\ref{fig:eitNonEitDominik}(a) and \ref{fig:eitNonEitDominik}(b), the parameters of the \textit{ab initio} level scheme description are given in Table~\ref{tab:symmetric_nondiagonal}. As expected from the discussion in Sec.~\ref{sec:micromacro}, the frequency shifts and coherent couplings between the two resonant layers $\Delta_{ll'}$ as well as the decay enhancements and incoherent couplings $\gamma_{ll'}$ are found to be identical for the two settings, since they depend on the electromagnetic environment only. 
However, when the cavity is illuminated from one side, the electronic mirror symmetry of the structure is broken for the incoupled light, and the Rabi frequencies driving the collective excitations differ in the two cavities. The reason for this is that in cavity Fig.~\ref{fig:eitNonEitDominik}(b), the light has to travel a larger distance to reach the resonant layers. Due to this additional half-wavelength propagation (with the effective wavenumber being reduced by the grazing incidence operation and the amplitude being modified by multiple reflections in the cavity), the field driving the off-center resonant layer acquires a relative sign flip as compared to that in the cavity in (a), while the absolute values of the driving fields in the two cavities are comparable. 
 
These additional relative phases then affect the matrix contractions in the spectra in Eqs.~\eqref{eq:reflection} and \eqref{eq:transmission}, and modify the interference of the different contributions, thus explaining the distinct appearance of the two reflection spectra. This interpretation can be further corroborated by considering the $\sigma^{-}_l$, which exhibit  deviating quantum dynamics in the two cases due to the different driving fields, even though the underlying multi-level schemes are equal. 

Since the couplings between the two excited states are equal in the EIT- and in the non-EIT cavity structures, the interpretation differs from previous ones based on the observable reflection spectrum, as already suggested in Sec.~\ref{sec:micromacro}. Within the interpretation in terms of the reflection spectrum, the vanishing EIT signatures were attributed to a negligible coupling between the involved excited states~\cite{rohlsberger_electromagnetically_2012}.
Using the diagonal form of the spectrum introduced in Sec.~\ref{sec:normalTheory}, we can understand this difference in detail. It allows us to quantify the contributions of both Lorentzians to the observable spectra (see Tab.~\ref{tab:symmetric_diagonal}), thus providing a clear explanation for their appearance. Since the coupling matrices in Table~\ref{tab:symmetric_nondiagonal} are identical for the two cavities, also the eigenvalues $\lambda$ are the same for both settings, see Table~\ref{tab:symmetric_diagonal}. In contrast, the spectral weights of the Lorentzians contributing to the reflection spectrum $g_l^{\mathrm{r}}$  are vastly different. For the cavity of Fig.~\ref{fig:eitNonEitDominik}(a), both eigenvalues contribute significantly to the reflection spectrum. However, for the case of Fig.~\ref{fig:eitNonEitDominik}(b), the contribution of the first, spectrally narrow Lorentzian to the reflection, and thereby the EIT feature, is almost completely suppressed. 

Contrary to that, for the transmission spectrum, the weights $g_l^\mathrm{t}$ coincide for both cavity settings (see Table~\ref{tab:symmetric_diagonal}). This explains the agreement of the two transmission spectra. While the driving of the excited states introduces an asymmetry between the cavity settings of Fig.~\ref{fig:eitNonEitDominik}(a) and \ref{fig:eitNonEitDominik}(b), the latter is compensated for by the outcoupling step, which is a direct consequence of the symmetry of the cavities. The transmission spectrum in Fig.~\ref{fig:eitNonEitDominik}(f) can be interpreted in terms of an underlying two-level system. Yet, this two-level interpretation is incompatible with the reflection spectra Fig.~\ref{fig:eitNonEitDominik}(d,e). 

Summarizing, the interpretation in terms of the reflection spectrum associates a multi-level system on the basis of the overall observable, including the particular in- and outcoupling contributions in the reflection geometry. The analogous interpretation in terms of the transmission spectrum may lead to a different multi-level interpretation, since the outcoupling and the electronic contribution differ from the reflection case. In contrast, the \textit{ab initio} approach directly provides a multi-level system describing the nuclear dynamics within the cavity, independent of the in- and outcoupling. Variations in the incoupling then translate into different Rabi frequencies driving the different transitions of the level scheme. The outcoupling only is of relevance in calculating observables.

This explains why an interpretation of the cavity systems in terms of observables may lead to inconsistent conclusions on the underlying level scheme, whereas the \textit{ab initio} description provides a unique level scheme, independent of the considered observables. In general, it differs from those obtained via the observable spectra, as illustrated by the coupling of the excited states in the EIT- and non-EIT case. 

However, it is important to stress that the desired application decides whether the underlying artificial level scheme or the observable spectra are in the focus of interest. For example, if a particular shaping of the x-ray light in frequency space is desired, then an interpretation and characterization in terms of the observable reflection spectrum may be most favorable. On the other hand, if, for example, effects of higher x-ray intensities on the interaction of the x-rays with the nuclei are to be considered, then an interpretation in terms of the observable spectra is likely to fail, since the nonlinearities affect the nuclear dynamics, but not the in- and outcoupling. Then, it appears more favorable to interpret the system in terms of the \textit{ab initio} approach, which separates the multi-level system from the in- and outcoupling. From the viewpoint of the inverse design, this difference also implies that optimization goals formulated in terms of the observable spectra may lead to different cavity structures than the corresponding optimization based on the parameters of the \textit{ab initio} multi-level scheme.

For this reason, both the inverse design of the \textit{ab initio} level scheme and that of the reflection spectrum are desirable. In the following, we discuss the two cases separately (Sec.~\ref{sec:invth} and Sec.~\ref{sec:reflection}).

\section{\label{sec:invth}Inverse design of artificial multi-level schemes within the cavity}
We start by considering the inverse design of artificial multi-level schemes based on the nuclear dynamics inside the cavity, using the \textit{ab initio} framework of~\cite{lentrodt_ab_2020}, and extending the analysis for two-level systems in~\cite{diekmann_inverse_2022}. To this end, we employ the example of thin-film cavities with two resonant layers, which realize artificial three-level schemes. We first introduce the three-level scheme description for this setup. Next, we provide the accessible level scheme properties, and subsequently show how one can design quantum optical schemes using the inverse approach.

\subsection{Artificial three-level systems}
\begin{figure}
\includegraphics[]{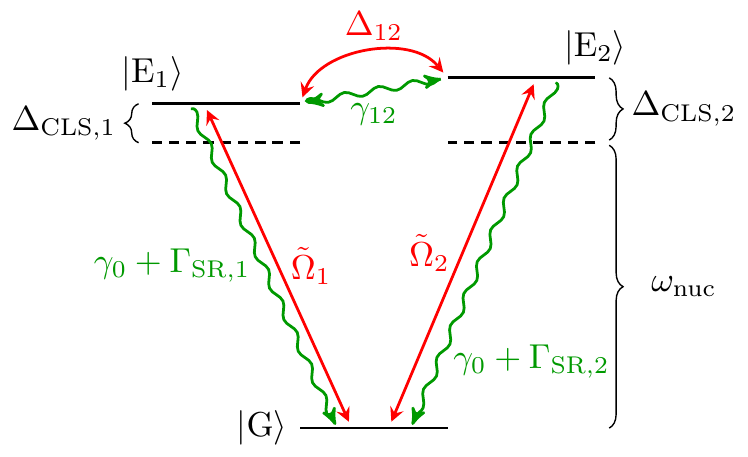}
\caption{Artificial nuclear three-level scheme as realized using a thin-film cavity with two resonant layers. The scheme comprises a ground state $\ket{\mathrm{G}}$ and two collective excited states $\ket{\mathrm{E}_{1(2)}}$. The transition frequency between ground and excited states is shifted from the individual nuclear transition frequency $\omega_{\mathrm{nuc}}$ by $\Delta_{\mathrm{CLS, 1(2)}}$ and the decay rates to the ground state are enhanced superradiantly by $\Gamma_{\mathrm{SR, 1(2)}}$ from the bare nuclear decay rate $\gamma_0$. Furthermore, the transitions are driven by the collectively enhanced Rabi frequencies $\tilde{\Omega}_{1(2)}$. Lastly, the thin-film cavity mediates a coherent coupling $\Delta_{12}$ as well as an incoherent cross-decay term $\gamma_{12}$ between the excited states. \label{fig:ThreeLevelNuclear}}%
\end{figure}

Specializing to two resonant layers placed in the cavity,  spin-waves can be excited in each of the layers, reducing the Hamiltonian Eq.~\eqref{eq:single1} to 
\begin{align} \label{eq:3level}
    \hat{H}=&\frac{\omega_{\mathrm{nuc}}}{2}\opsigma^z_{1}+\Delta_{\mathrm{CLS,1}}{\opsigma}^+_{1}{\opsigma}^-_{1}+\frac{\omega_{\mathrm{nuc}}}{2}\opsigma^z_{2}\notag\\&+\Delta_{\mathrm{CLS},2}{\opsigma}^+_{2}{\opsigma}^-_{2}-\left(\Delta_{12}{\opsigma}^+_{1}{\opsigma}^-_{2}+\mathrm{H.c.}\right)\notag\\
    &-\left(\Omega_1{\opsigma}^+_{1}+\mathrm{H.c.}\right)-\left(\Omega_2{\opsigma}^+_{2}+\mathrm{H.c.}\right)\,.
\end{align} 
Here, we introduced the frequency shifts [c.f. Eq.~\eqref{eq:DeltaEff}]
\begin{align}
\Delta_{\mathrm{CLS},1}=-\Delta_{11}\,, \quad\Delta_{\mathrm{CLS},2}=-\Delta_{22}\,,
\end{align}
also known as collective Lamb shifts. Furthermore, the Hamiltonian comprises a dipole-dipole coupling of strength $\Delta_{12}$ between the excited states, where  $\Delta_{12}=\Delta_{21}$ is real and symmetric.

According to Eq.~\eqref{eq:GammaEff}, the associated Lindbladian describes superradiant decay enhancements of the excited states, 
\begin{align}
\Gamma_{\mathrm{SR},1} = \gamma_{11}\,,\quad\Gamma_{\mathrm{SR}, 2}=\gamma_{22}\,
\end{align} 
as well as an incoherent cross-coupling term $\gamma_{12}$, where $\gamma_{12}=\gamma_{21}$ is also real and symmetric. Such cross-decay terms are typically not present in naturally occurring three-level systems such as atoms, but have been intensely studied in the literature as they are known to give rise to so-called vacuum-induced or spontaneously generated coherences, which form the basis of a multitude of applications~\cite{hohler_quantum_1974,ficek_quantum_2005,kiffner_vacuum-induced_2010,heeg_vacuum-assisted_2013}.

In principle, the above Hamiltonian Eq.~(\ref{eq:3level}) describes the effective dynamics in a four-dimensional space. The Hamiltonian is, however, derived in the experimentally relevant low-excitation regime, $\sigma^z_j\approx -1$. Here, only the ground state, $\ket{\mathrm{G}}\equiv\ket{\mathrm{g}_1\mathrm{g}_2}$, and singly excited states, $\ket{\mathrm{E}_1}\equiv\ket{\mathrm{e}_1\mathrm{g}_2}$, $\ket{\mathrm{E}_2}\equiv\ket{\mathrm{g}_1\mathrm{e}_2}$, contribute to the dynamics, thus leaving us with a three-level description. 

\subsection{Multi-level-scheme parameters}\label{sec:invth:bare}
For concreteness, in the following discussion, we consider  an experimentally relevant thin-film cavity of the type shown in Fig.~\ref{fig:interrelations} deposited on a silicon substrate with the material combination (top to bottom) Pd/\allowbreak \ch{B4C}\allowbreak /\Fe{}/\allowbreak\ch{B4C}/\Fe{}/\allowbreak\ch{B4C}/\allowbreak Pd/\allowbreak Si. Here, Pd represents the cladding material and \ch{B4C} the guiding material. The resonant \Fe layers are assumed to have a thickness of $0.574~\mathrm{nm}$, corresponding to about two layers of \Fe nuclei. Note that one may also consider, e.g., an additional mirror layer in between the two iron layers~\cite{haber_rabi_2017}, however, such cavities can be treated using analogous methods. 

The three-level system parameters resulting from the thin-film structure with two resonant layers depend on the various layer thicknesses and the angle of incidence. In a first step, we determine the experimentally accessible range of quantum optical level schemes. To that end, we employ the analytically known Green's function of the cavity~\cite{tomas_green_1995} together with the numerical methodology introduced in~\cite{diekmann_inverse_2022}. To keep the notation clear and concise, we refer to the different quantum optical parameters of the three-level scheme as \textit{observables} and call the set of all accessible values for a combination of observables under study an \textit{observables space} (OS). Since we compare different combinations of observables, we refer to them as different observables spaces (OSs). The observables are functions of the geometrical parameters of the thin-film structure, i.e., the layer thicknesses and the angle of incidence. We refer to the latter as \textit{cavity parameters}.

\subsubsection{\label{sec:flp}Coupling rates and complex level shifts}\label{sec:invth:designCoupling}
\begin{figure*}
\includegraphics{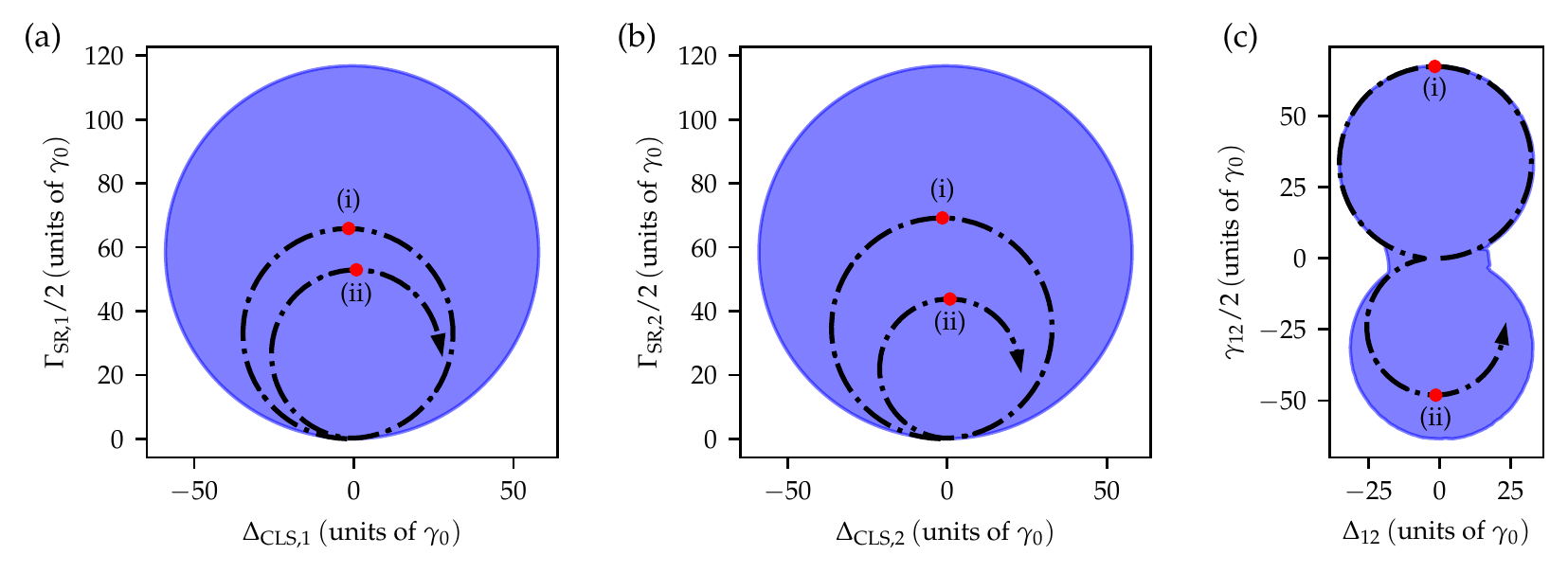}
\caption{Observable spaces characterizing the artificial three-level system implemented using a Pd/\ch{B4C}/\Fe/\ch{B4C}/\allowbreak\Fe/\allowbreak\ch{B4C}/\allowbreak Pd/\allowbreak Si cavity setup. (a,b) the accessible CLS and SR rates of the two excited states are shown in blue, respectively. (c) shows the accessible cavity-mediated coherent and inherent coupling rates between the two excited states. 
The black dash-dotted lines represent trajectories  that are obtained upon changing the angle of incidence for  the specific cavity of highest incoherent coupling, described in Fig.~\ref{fig:PoleStructureCouplings}. The arrow denotes the direction of increasing $\theta$. Red dots on the trajectories, labeled with lower-case roman numerals,
relate to the cavity resonances in Fig.~\ref{fig:PoleStructureCouplings}. Note, that the range of angles used for the trajectories is smaller than in Fig.~\ref{fig:PoleStructureCouplings}, for better readability. However, the angular range is the same for all panels of the present plot.}
\label{fig:ThreeLevelSysBareCouplings}
\end{figure*}

The structure of the level scheme is defined via the frequency shifts of both excited states and their respective decay enhancements (which together form a complex level shift), as well as the coherent and incoherent couplings of the excited states. The corresponding OSs for these observables are shown in Fig.~\ref{fig:ThreeLevelSysBareCouplings}. In the following, we first discuss the OSs and then explain their appearance on the basis of the cavity mode structure.

\begin{figure}
\includegraphics[]{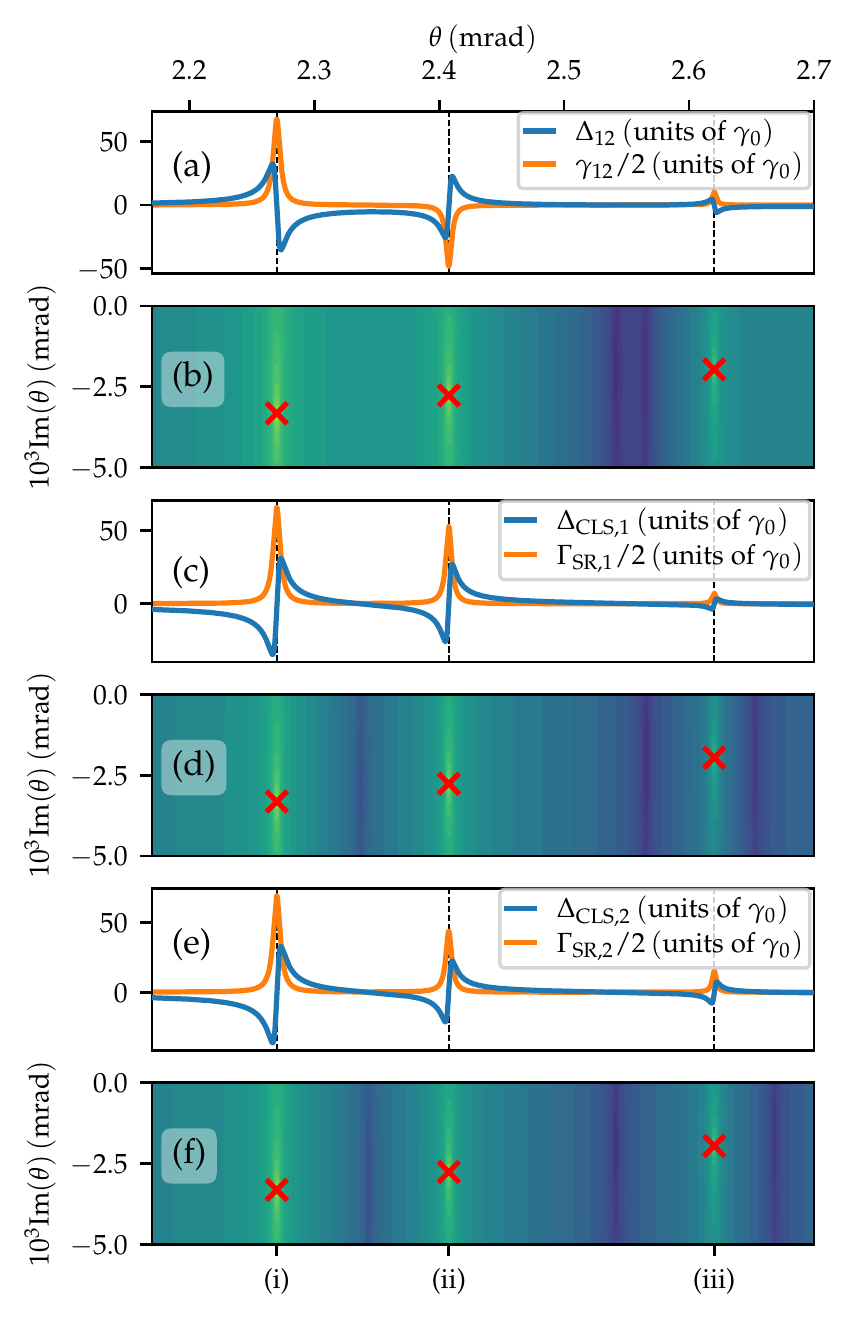}
\caption{Analysis of the cavity with highest incoherent coupling rate $\gamma_{12}$ in Fig.~\ref{fig:ThreeLevelSysBareCouplings} as a function of the x-ray incidence angle $\theta$. Panel (a) shows the coherent and incoherent coupling rates between the two excited states, whereas (c,e) show the CLS and the SR of the two excited states, respectively. Panels (b,d,f) show the absolute values of the complex coupling rates $\Delta_{12}+i\gamma_{12}/2$, $\Delta_{\textrm{CLS},1} + i\Gamma_{\textrm{SR},1}/2$ and $\Delta_{\textrm{CLS},2} + i\Gamma_{\textrm{SR},2}/2$ in the complex $\theta$ plane, respectively. The poles corresponding to the resonances in the couplings are indicated by red crosses. The black dashed lines in the plots indicate the real parts of the respective poles. The latter are enumerated by roman numerals and relate the poles to the corresponding circular trajectories in Fig.~\ref{fig:ThreeLevelSysBareCouplings}. All poles coincide as the different multi-level system parameters  relate to the same underlying Green's function.
The cavity layout is Pd($105.1\,\mathrm{nm}$)/\ch{B4C}($27.7\,\mathrm{nm}$)/\Fe{}($0.57\,\mathrm{nm}$)/\ch{B4C}($23.8\,\mathrm{nm}$)/\allowbreak\Fe{}($0.57\,\mathrm{nm}$)/\allowbreak\ch{B4C}($28.8\,\mathrm{nm}$)/\allowbreak Pd($12.5\,\mathrm{nm}$)/\allowbreak Si. 
}
\label{fig:PoleStructureCouplings}
\end{figure}

From Figs.~\ref{fig:ThreeLevelSysBareCouplings}(a) and \ref{fig:ThreeLevelSysBareCouplings}(b), we see that the OSs relating the accessible respective level shifts and decay rate enhancements of the two individual transitions assume circular shapes. Analogous observations were reported in~\cite{diekmann_inverse_2022} for the single transition in cavities with one resonant layer. 
In contrast, the corresponding OS relating the accessible coherent and incoherent coupling parameters between the  two excited states has a different structure [see Fig.~\ref{fig:ThreeLevelSysBareCouplings}(c)]. It essentially consists of two circles, one of which is oriented towards positive values for the incoherent couplings $\gamma_{12}$, while the other one is oriented towards negative values. 

To understand these OS shapes, we focus on the cavity geometry leading to the highest possible incoherent coupling in Fig.~\ref{fig:ThreeLevelSysBareCouplings}(c). Upon varying only the angle of incidence, the resulting observables traverse the black dash-dotted trajectories indicated in Figs.~\ref{fig:ThreeLevelSysBareCouplings}(a,b,c). Along these trajectories, the roman numerals indicate the different cavity resonances.

For a more detailed analysis, we show the observables along the above trajectories as a function of $\theta$ in Figs.~\ref{fig:PoleStructureCouplings}(a,c,e). Further, we plot the absolute values of the corresponding three complex-valued couplings (formed as $\Delta  + i\gamma/2 $ from the corresponding two real-valued parameters) in the complex $\theta$ plane in Figs.~\ref{fig:PoleStructureCouplings}(b,d,f) to relate the observables of the three-level scheme to their poles as a function of $\theta$. These poles can be associated to the mode structure of the cavity~\cite{lalanne_light_2018, diekmann_inverse_2022, lentrodt_classifying_2021}. To explain the circularly-shaped trajectories, the Green's function can be expressed using a Mittag-Leffler pole expansion~\cite{diekmann_inverse_2022,lalanne_light_2018, arfken_mathematical_2013},
\begin{align}
{G}(z_l, z_{l'}, \vec{k}_\parallel(\theta) , \omega_\mathrm{nuc}) = {G}(z_l, z_{l'}, \vec{k}_\parallel(\theta=0) , \omega_\mathrm{nuc})\notag\\ + \sum\limits_{\theta_0}\mathrm{Res}(G, \theta_0)\left(\frac{1}{\theta_0}+\frac{1}{\theta-\theta_0}\right)\,,
\label{eq:MLGreen}
\end{align}
with the poles $\theta_0$ of the Green's function evaluated as a function of $\theta$. Here, $\Res(G, \theta_0)$ denotes the residue of the Green's function at $\theta_0$. For a single pole $\theta'$ well-separated from neighboring ones, the Green's function can be approximated by a single mode expression,
\begin{equation}
G(z_l, z_{l'}, \vec{k}_\parallel(\theta) , \omega_\mathrm{nuc}) \approx \mathcal{C}+\frac{\mathrm{Res}(G, \theta')}{\theta-\theta'}\,,
\label{eq:ML_SM}
\end{equation}
where the relevant terms constant in $\theta$ are summarized in $\mathcal{C}$. As a result, we find that   circular trajectories in the complex parameter plane indeed are to be expected, as long as the single mode approximation is applicable~\cite{rassias_circle_2014,diekmann_inverse_2022}. Since all observables considered in Figs.~\ref{fig:ThreeLevelSysBareCouplings} and \ref{fig:PoleStructureCouplings} involve a single Green's function, they share the same poles as a function of $\theta$, as illustrated in Fig.~\ref{fig:PoleStructureCouplings}. However, it is important to note that they may differ in their corresponding residues.

Upon passing the first pole [indicated by (i)] on the real $\theta$-axis, an (approximate) circle is analogously traversed for all three observables pairs in Fig.~\ref{fig:ThreeLevelSysBareCouplings}. However, the situation is different for the second pole: Here, the couplings between the excited states in (c) traverse a downward-oriented circle involving negative $\gamma_{12}$,  in contrast to the two other observable pairs in panels (a,b). Using the single mode approximation Eq.~\eqref{eq:ML_SM}, this can be explained by a sign switch of the residues of neighboring poles for the couplings between the excited states, which is not present  for the superradiant decay rate enhancements and frequency shifts. Another difference is  that   Fig.~\ref{fig:ThreeLevelSysBareCouplings}(c)  also shows a small region of accessible observables in between the two circular structures, close to $\gamma_{12}=0$. This region can be associated to very small angles of incidence. 

A particular trait of the trajectory in Fig.~\ref{fig:ThreeLevelSysBareCouplings}(c) is that it partly runs on the surface of the OS. Therefore, a variety of extremal couplings between the excited states is found in a single cavity geometry. Note that analogous circular structures upon variation of the incidence angle were already observed previously for the decay enhancement and level shift of individual transitions~\cite{longo_tailoring_2016, diekmann_inverse_2022}.

Since all observables share the same poles of the Green's function, their dependencies on the angle of incidence are related. As a result, the observables cannot be tuned  independently from each other in any given cavity structure. Our systematic analysis of the accessible observables nevertheless shows that a large variety of parameter combinations can be realized via a suitable choice of the structure. 

\begin{figure}
\includegraphics[]{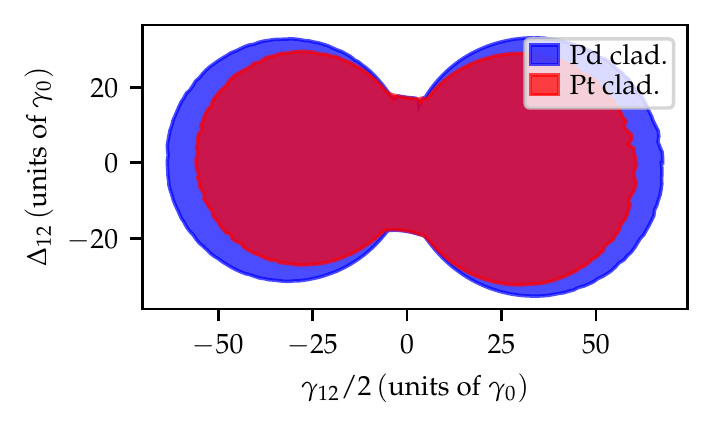}
\caption{Dependence of the artificial multi-level system tunability on the cavity materials. The figure compares the OSs of the accessible  coherent and incoherent couplings for Pt and Pd as cladding materials in a cladding/\ch{B4C}/\Fe/\ch{B4C}/\allowbreak\Fe/\allowbreak\ch{B4C}/\allowbreak cladding/\allowbreak Si cavity setup. The less absorptive Pd allows for larger tunability than Pt cladding layers in accordance with observations in cavities with single resonant layer~\cite{diekmann_inverse_2022}. \label{fig:DoubleCheese_PdVsPt}}%
\end{figure}

To conclude the discussion of the multi-level system parameters, we briefly discuss their dependence on the cavity materials. As an example, Fig.~\ref{fig:DoubleCheese_PdVsPt} compares the accessible couplings between the excited states for the cavity in Fig.~\ref{fig:ThreeLevelSysBareCouplings} with a corresponding cavity in which the mirror material Pd has been replaced by Pt.  It can be seen that Pd outperforms Pt as mirror material in terms of the parameter tunability. Interestingly, Pd is less refractive ($\delta_\mathrm{Pd}<\delta_\mathrm{Pt}$, where the refractive index is $n = 1 -\delta + i \beta$), but also less absorptive ($\beta_\mathrm{Pd}<\beta_{Pt}$). This result therefore is consistent with one of the general design guidelines derived for single-layer cavities in~\cite{diekmann_inverse_2022}, i.e., the performance of the cavities is primarily determined by the absorptive properties of the involved materials. It therefore suggests that the significance of low absorption in the cladding layer also translates to systems with multiple resonant layers.

\subsubsection{\label{sec:intra}Rabi frequencies}
Besides the multi-level system parameters discussed in the preceding Sec.~\ref{sec:flp}, also the Rabi frequencies $\tilde{\Omega}_1$ and $\tilde{\Omega}_2$ describing the coupling of the externally applied x-rays to the level scheme crucially influence the nuclear dynamics, and thus the system properties. These Rabi frequencies depend on the externally applied x-ray field inside the cavity. For the cavity used in Fig.~\ref{fig:PoleStructureCouplings}, the resulting driving fields are negligibly small, as the thick top cladding layer effectively prevents any outside field from entering the cavity. The reason is that -- as already observed in the case of a single layer~\cite{diekmann_inverse_2022} -- the design goals of high intra-cavity field enhancements and of large magnitudes of the multi-level parameters are   incompatible in general. Therefore, in the following, we discuss the Rabi frequencies separately.

The Rabi frequencies $\tilde{\Omega}_1$ and $\tilde{\Omega}_2$ are proportional to the field strength (in-plane Fourier transformed and in the frequency domain) at the respective layer, ${E}_\mathrm{in}(z_l, \vec{k}_\parallel, \omega_\mathrm{nuc})$.  To arrive at a comprehensive description of the three-level scheme observables, we therefore determine the OSs of accessible field strengths, ${E}_\mathrm{in}(z_1, \vec{k}_\parallel, \omega_\mathrm{nuc})$ and ${E}_\mathrm{in}(z_2, \vec{k}_\parallel, \omega_\mathrm{nuc})$, at the respective resonant layers. We consider the accessible real and imaginary parts of the driving fields to illustrate similarities of the field enhancements' behaviors to the remaining level scheme properties. First, we consider the accessible OSs for the field enhancements. Subsequently, we again specialize to one cavity geometry and tune the angle of incidence to explain the structure of the OSs found on the basis of the poles of the field enhancements.

\begin{figure*}
\centering
\includegraphics[width=0.95\textwidth]{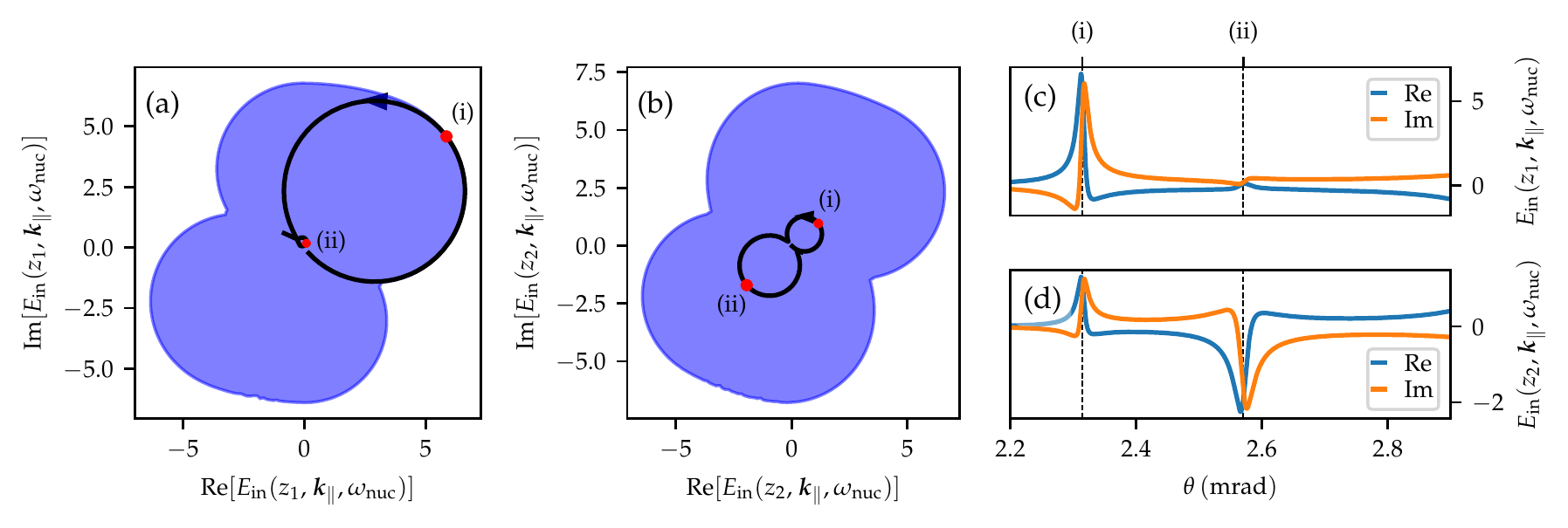}
\caption{Observable spaces characterizing the set of accessible values for the x-ray field enhancements at the resonant layers. The field enhancements determine the respective Rabi frequencies of the multi-level system. The blue shaded areas in (a) and (b) show the accessible real and imaginary parts of the field strength at the first and second nuclear layer for Pd/\ch{B4C}/\Fe/\ch{B4C}/\allowbreak\Fe/\allowbreak\ch{B4C}/\allowbreak Pd/\allowbreak Si cavity setups, respectively. The black lines in (a) and (b) correspond to the trajectories traversed upon tuning the angle of incidence $\theta$ for the Pd($3.8\,\mathrm{nm}$)/\ch{B4C}($27.1\,\mathrm{nm}$)/\allowbreak\Fe{}($0.57\,\mathrm{nm}$)/\ch{B4C}($27.6\,\mathrm{nm}$)/\allowbreak\Fe{}($0.57\,\mathrm{nm}$)/\allowbreak\ch{B4C}($0\,\mathrm{nm}$)/\allowbreak Pd($41.6\,\mathrm{nm}$)/\allowbreak Si cavity of highest real part field enhancement in the first resonant layer. Here, the arrows indicate increasing $\theta$. For this cavity, the explicit dependency of the field enhancement in the first resonant layer on $\theta$  is shown in panel (c) and the real parts of its poles in the complex $\theta$ plane are indicated as dashed lines. Likewise, panel (d) shows the analogous plots for the second resonant layer. Lastly, the red dots with annotated roman numerals link the circular structures in (a) and (b) to the poles of the respective quantities. The trajectories with the angle of incidence are reminiscent of the ones in Fig.~\ref{fig:ThreeLevelSysBareCouplings}. Together with the common pole structure of all observables, this emphasizes the intertwined dependencies of the observables on some cavity parameters.}
\label{fig:Double_Cheese_bareCouplings_fields}
\end{figure*}

The resulting OSs for the field enhancements are shown in Figs.~\ref{fig:Double_Cheese_bareCouplings_fields}(a) and (b). The OS for the real and imaginary field enhancement in the first resonant layer is found to be similar, but not identical, to the one in the second resonant layer. Overall, the structure of both OSs shares some similarity with the corresponding OS of the coupling between the excited states in Fig.~\ref{fig:ThreeLevelSysBareCouplings}(c). 

Similar to the discussion of the trajectories in Fig.~\ref{fig:ThreeLevelSysBareCouplings}, we focus on the specific cavity of highest real field enhancement at the first nuclear layer, and study its observables as a function of the angle of incidence. The resulting trajectories are plotted as black curves in Figs.~\ref{fig:Double_Cheese_bareCouplings_fields}(a,b). Panels (c,d) further show the resulting real and imaginary parts of the field enhancements as a function of the angle of incidence. Note that since the field enhancements also rely on the same underlying Green's function governed by the cavity structure, they share a common pole structure with the remaining coupling constants.

From Fig.~\ref{fig:Double_Cheese_bareCouplings_fields}(a) and (b) it is apparent that also the field strengths traverse circular trajectories when $\theta$ passes by a respective pole, indicated by (i) or (ii). However, the trajectory of highest real field enhancement in the first nuclear layer corresponds to rather low field enhancements at the second layer. Furthermore, as compared to Fig.~\ref{fig:ThreeLevelSysBareCouplings}, the circles are found to be more pronouncedly rotated in the complex plane. This can be attributed to the fact that the residues of the field enhancements are complex-valued, thus leading to different orientations of the single-mode circles in the complex plane. As an important consequence,  the accessible OSs obey different symmetries as compared to those found for the remaining observables of the level scheme in Fig.~\ref{fig:ThreeLevelSysBareCouplings}, which allows for a more independent tuning of the Rabi frequencies from the other coupling constants.

This concludes our basic discussion of the OSs for the relevant properties of the three-level scheme. Since the approach allows one to access the cavity parameters for the samples calculated in the OS, the inverse design can straightforwardly be  applied to systems with two resonant layers. Note that due to the higher number of level scheme properties involved, not all relations between relevant observables were shown explicitly. This is important, since the different observables are correlated such that, e.g., particular desired values for one observable may not be simultaneously realizable with corresponding choices for another observable. Using the formalism of~\cite{diekmann_inverse_2022}, however, the OSs of arbitrary observable combinations can be readily calculated. 

In the following, we extend the discussion to ratios of three-level scheme observables, e.g., $\Delta_{12}/(\Gamma_{\mathrm{SR},1}+\gamma_0)$. Such ratios in general are crucial for the interpretation in terms of the quantum optical functionality, as the relevant timescale of the dynamics in the three-level system usually is set by the decay rates of the excited states.

\subsection{\label{sec:invth:relative}Multi-level-scheme parameter ratios}

In the previous Sec.~\ref{sec:intra}, we discussed the relevant properties of the artificial three-level system. In this context, we have seen that the different observables determining the effective level scheme cannot be chosen independently from one another, because of their dependence on the same underlying Green's function determined by the cavity structure. From a quantum optical perspective, however, the qualitative nature of the dynamics typically is not governed by the absolute properties of the level scheme, but by their ratios. 

Of particular importance are  the decay times on the different transitions, since they determine the relevant timescales of the overall dynamics in the considered low-excitation regime. For example, the effect of the cavity-mediated couplings between the excited states depends on their ratios to the corresponding decay rates. At the same time, these cavity-mediated couplings between excited states are crucial for the design of advanced quantum optical schemes, because they may replace otherwise unavailable coherent external x-ray driving fields.
Because of their significance, the ratios of the coupling constants to the respective decay rates will be in the focus of the following discussion.

Since the effective three-level scheme comprises two excited states $\ket{\mathrm{E}_{1}}, \ket{\mathrm{E}_{2}}$, it is not a priori clear which of the two state decay rates determines the relevant timescale. For this reason, we will start the discussion in Sec.~\ref{sec:scaled} by considering the coherent and incoherent couplings scaled by the first ($\Gamma_{\mathrm{SR,1}}+\gamma_0$) and second ($\Gamma_{\mathrm{SR,2}}+\gamma_0$) excited state decay rate separately. Subsequently, in Sec.~\ref{sec:three}, we will complement the results by adding the remaining decay rate as a third observable.

\subsubsection{\label{sec:scaled}Scaled couplings between excited states}

\begin{figure*}
\includegraphics{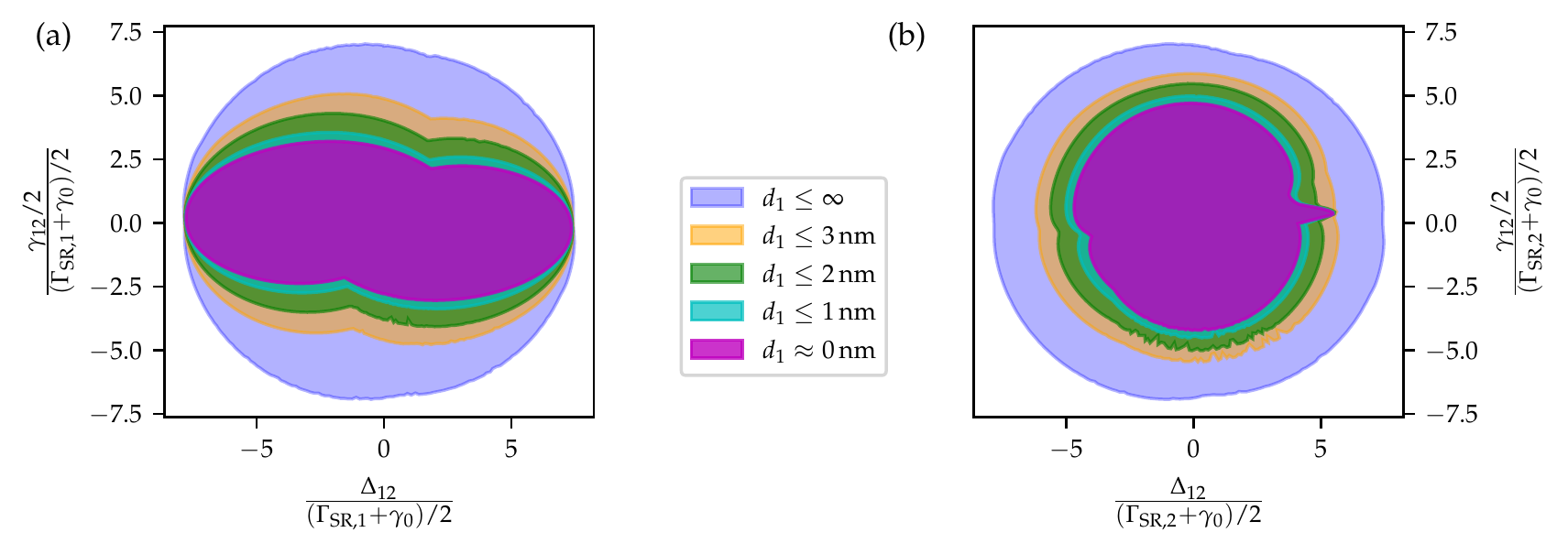}
\caption{Accessible OSs for the excited states' coherent and incoherent couplings scaled by the decay rate of the first (a) and second (b) excited state. To ensure the incoupling of the driving field and the outcoupling of the nuclear signal, different maximal thicknesses for the top cladding layer $d_1$ %(see Fig.~\ref{fig:schematicSetupThree}(a)) 
are considered, and shown as the shaded areas. The results were obtained for a Pd/\ch{B4C}/\Fe/\ch{B4C}/\allowbreak\Fe/\allowbreak\ch{B4C}/\allowbreak Pd/\allowbreak Si cavity setup.  \label{fig:doubleCheese_couplings_scaledGamma}}
\end{figure*}

Results for the accessible OSs for the coherent and incoherent couplings scaled by the first and second decay rates, respectively, are shown in Fig.~\ref{fig:doubleCheese_couplings_scaledGamma}. In these panels, the different shaded areas correspond to various constraints on the top mirror thickness.  

Without constraints on the top mirror layer thickness ($0\leq d_1<\infty$), we find that the OSs exhibit roughly circular shapes. Moreover, the two OSs in (a,b) essentially agree with each other. This can be understood as follows. In principle, the two resonant layers are embedded in different photonic environments, as the upper layer is closer to the top surface, i.e., vacuum, while the lower layer is closer to the Si substrate. Generally, the OSs for the scaled couplings should therefore be expected to be different depending on which decay rate is used for scaling. However, the largest coupling rates are usually realized for very thick cladding layers. Such thick claddings suppress the effect of the asymmetry induced between air and substrate on either side of the cavity structure, such that without constraints of the top layer thickness, both OSs nevertheless are similar.

Since for thick cladding layers the coupling into and out of the cavity typically is small, the corresponding signatures of the level scheme in the observable spectra generally are highly suppressed. In order to ensure an observable imprint on the spectra, and thus the  relevance for experiments measuring the reflected light, we subsequently impose the additional constraint of restricting  the top cladding layer to maximum thicknesses of $3\,\mathrm{nm}$, $2\,\mathrm{nm}$, $1\,\mathrm{nm}$ as well as to no cladding layer at all. The resulting OSs are shown by the different shaded areas in Fig.~\ref{fig:doubleCheese_couplings_scaledGamma}.

As expected, the additional constraints reduce the tunability of the achievable scaled coupling strengths. Furthermore, the constraints also introduce a pronounced asymmetry between OSs in panels (a) and (b), in contrast to the result without constraint. This is because for lower cladding thicknesses, the asymmetric environment of the cavities becomes important.

Interestingly, in Fig.~\ref{fig:doubleCheese_couplings_scaledGamma}(a) we find that near-maximum absolute values of the scaled coherent coupling $\Delta_{12}$ can be realized independent of the upper cladding layer thickness, and even without an upper cladding layer. However, the range of the accessible incoherent coupling rates shrinks with the maximum allowed top-layer thickness.

In the coupling rates scaled to the decay rate of the second transition in Fig.~\ref{fig:doubleCheese_couplings_scaledGamma}(b), a peculiar tip appears towards the maximum possible positive scaled coherent couplings when the top cladding layer is restricted to thicknesses below $2\,\mathrm{nm}$. Interestingly, this tip is preserved upon further reducing the thickness, and even for the case of a vanishing top cladding layer. As a result, also this  OS  allows for comparatively large scaled coherent couplings with cavities featuring a low top cladding thickness.

Note that it was already found previously for thin-film cavities with one resonant layer that cavities without top mirror can form relevant and interesting alternatives to traditional cavity designs~\cite{diekmann_inverse_2022}.
 
\subsubsection{\label{sec:three}Scaled couplings vs. decay rate ratios} 
\begin{figure} 
\includegraphics[width=0.9\columnwidth]{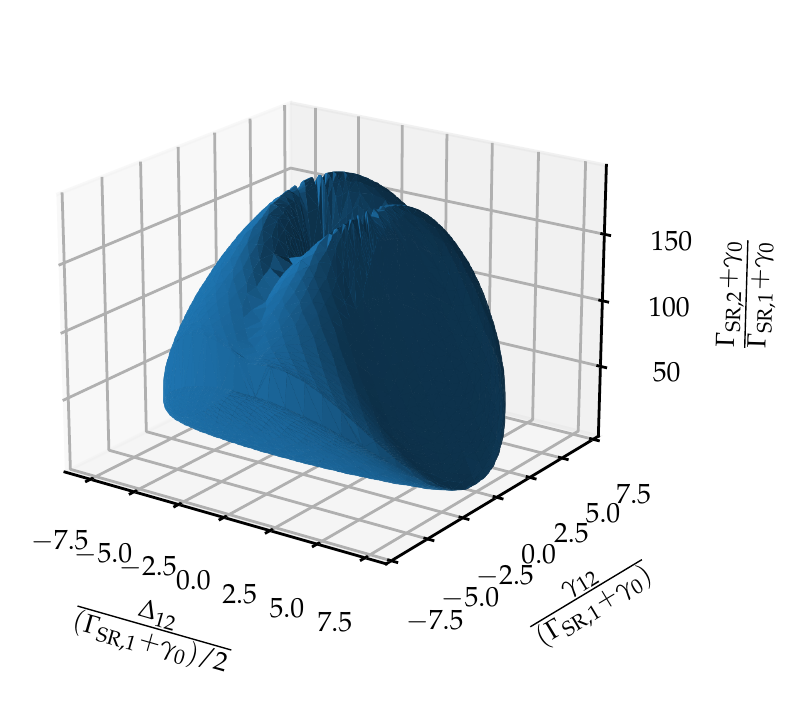}
\caption{Accessible OS for the excited state coherent and incoherent couplings combined with the decay rate of the second excited state. All quantities are scaled by the first excited state decay rate. The OSs were calculated for a Pd/\ch{B4C}/\Fe/\ch{B4C}/\allowbreak\Fe/\allowbreak\ch{B4C}/\allowbreak Pd/\allowbreak Si cavity setup. The same OS is found when the first and second excited state decay rate are exchanged.%As the topology of the surfaces of the OSs (a) and (b) is complex, we only give the surface points calculated with the methods of~\cite{diekmann_inverse_2022}, rather than trying to plot an explicit surface. We support the readability by projecting the points to the axes planes. 
\label{fig:DoubleCheese_couplings_gamma_scaledGamma}}
\end{figure}

So far, we considered the coupling between both excited states scaled by each of the decay rates independently. However, to fix the relevant time scale governing the system dynamics, the ratio between the decay rates of the 
two excited states has to be considered as well. Hence, we now consider the coherent and incoherent coupling between the excited states together with one of the excited state decay rates as a third observable, and scale all three rates by the remaining excited state decay rate. Fig.~\ref{fig:DoubleCheese_couplings_gamma_scaledGamma} shows the OS surface for the coherent and incoherent coupling and the second excited state decay rate, scaled by the first excited state decay rate. Note that as in Fig.~\ref{fig:doubleCheese_couplings_scaledGamma}, the corresponding calculation with the two decay rates interchanged results in a very similar surface (not shown), because we did not restrict the top cladding layer thickness. Overall, we find that a large and diverse set of observables combinations is offered by the artificial three-level system, despite the fact that all couplings derive from the same underlying Green's function. The two excited state decay rates can assume ratios larger than 100, while still offering a tunability of the excited state coupling rate. 

\begin{figure*}
\includegraphics[]{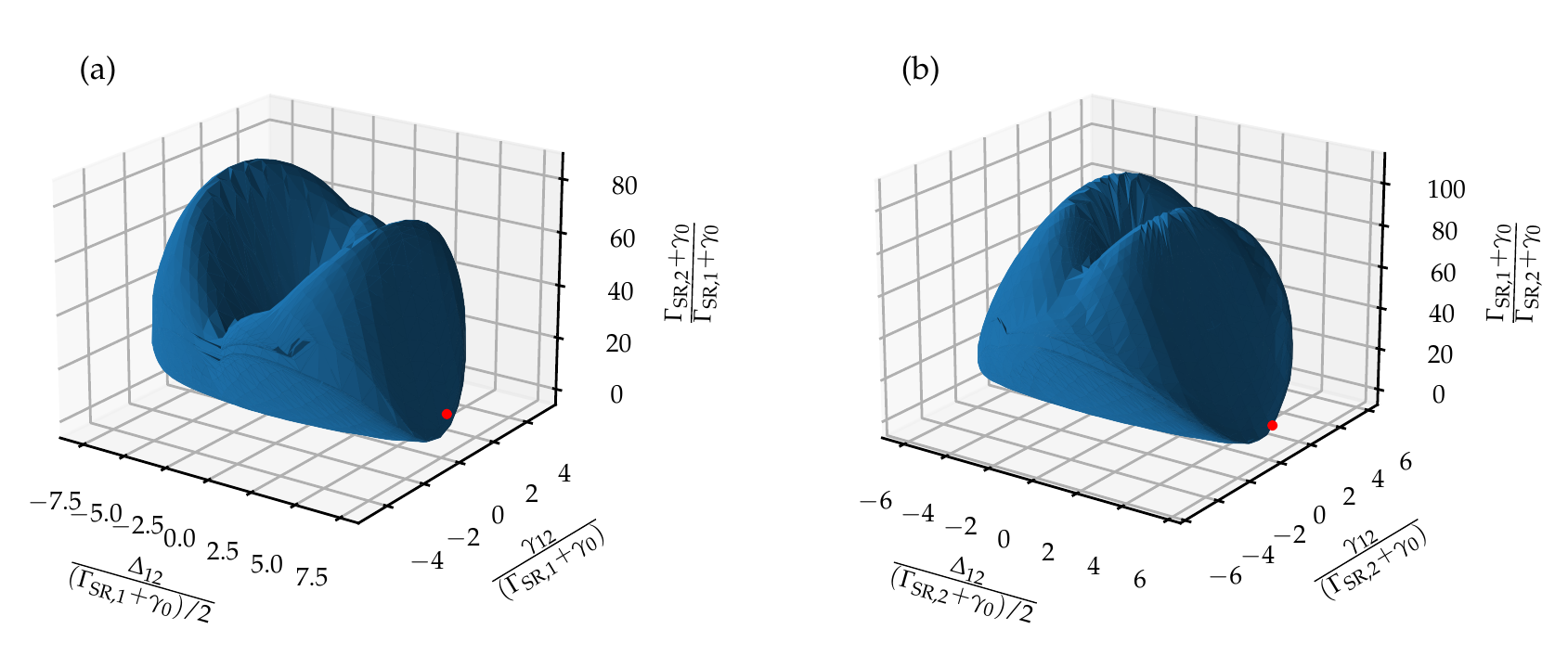}
\caption{Accessible OSs analogous to Fig.~\ref{fig:DoubleCheese_couplings_gamma_scaledGamma} but for the top cladding layer bounded by  $d_1\leq 3\,\mathrm{nm}$. In panel (a) and (b), the cavities corresponding to Tables \ref{tab:EIT} and \ref{tab:NoEIT} are indicated by red dots, respectively. %We only plot surface points analogously to Fig.~\ref{fig:DoubleCheese_couplings_gamma_scaledGamma}.  
\label{fig:DoubleCheese_couplings_gamma_scaledGamma_restrictedCladding}}
\end{figure*}

In order to ensure the level schemes' observability in experiments, we next consider the corresponding plots with top cladding layer thicknesses restricted to $\leq 3\,\mathrm{nm}$ (see Fig.~\ref{fig:DoubleCheese_couplings_gamma_scaledGamma_restrictedCladding}). As was already noticed for the two-dimensional version of the plots in Fig.~\ref{fig:doubleCheese_couplings_scaledGamma}, the restriction of the top cladding layer thickness leads to an asymmetry between the two OSs, which are derived by scaling with either of the two excited state decay rates.  Despite the additional constraint, a considerable tunability of the three-level system is still possible. Interestingly, the highest decay rate ratios tend to go along with quite large excited state couplings. However, the highest possible absolute values for the scaled coherent coupling are only realizable with low decay rate ratios.

%Since each observable chosen for the calculation is a ratio of three-level scheme observables, the respective OSs turn out not to be smooth surfaces, but show more complex topology in certain regions. We therefore restrict ourselves to plotting the surface points calculated and omit the interpolation by a smooth surface. For clarity, the projections to the axes planes are indicated. For the projection to the bottom see Fig.~\ref{fig:doubleCheese_couplings_scaledGamma}.

%Overall, Figs.~\ref{fig:DoubleCheese_couplings_gamma_scaledGamma}(a) and \ref{fig:DoubleCheese_couplings_gamma_scaledGamma}(b) show that a large set of observables combinations is accessible for the design. As previously, the two surfaces are found to be very similar for the reason that we did not restrict the top cladding layer thickness. However, this again has to be considered a problem for the driving and the observation of the artificial three-level scheme. 

OSs as shown in Fig.~\ref{fig:DoubleCheese_couplings_gamma_scaledGamma_restrictedCladding} provide the basis for the design of nuclear quantum optical effects that rely on the coherent and incoherent coupling between the excited states of the three-level scheme. In the following Sec.~\ref{sec:invth:EIT}, we will show how such OSs can be employed in practice to design a particular nuclear quantum optical scheme.

\subsection{\label{sec:invth:EIT}Inverse design of nuclear quantum optical effects -- electromagnetically induced transparency}

In this Section,  we apply the results obtained so far to the inverse design of a quantum optical effect. As a concrete example, we address the effect of EIT. An EIT-like behavior has been observed and interpreted in terms of the observable reflection spectrum in \cite{rohlsberger_electromagnetically_2012}. In contrast, here, we will focus on the design of the corresponding \textit{ab initio} multi-level scheme. This approach is complementary to that based on the observable spectra, as discussed in Secs.~\ref{sec:Introduction} and \ref{sec:micromacro}.

In the following, we first briefly revisit the basic concept of EIT, to provide a guideline for choosing the design goals. We subsequently engineer nuclear EIT schemes using the artificial three-level system realized in cavities with two resonant layers.
 
\subsubsection{EIT design criteria}\label{sec:genericEIT}
\begin{figure}
\includegraphics[]{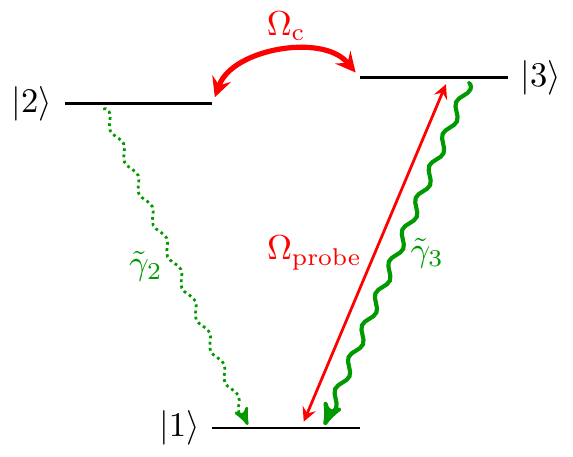}
\caption{Archetype three-level setup featuring electromagnetically induced transparency~\cite{rohlsberger_electromagnetically_2012}. Two states $\ket{1}, \ket{2}$ are assumed to be (meta)stable, with only a small residual decay rate $\tilde \gamma_2$. A control field $\Omega_c$ couples $\ket{2}$ to the excited state $\ket{3}$, which decays with comparably larger decay rate $\tilde\gamma_3$ to the ground state $\ket{1}$. Upon weakly probing the $\ket{1}\leftrightarrow\ket{3}$ transition with laser field $\Omega_\mathrm{probe}$, the EIT effect can be expected subject to the condition Eq.~\eqref{eq:EITcondition}.\label{fig:VLevelScheme}}%
\end{figure}

The archetypal quantum optical EIT scheme is realized in suitable three-level systems~\cite{fleischhauer_electromagnetically_2005}, e.g., the one shown in Fig.~\ref{fig:VLevelScheme}.  To observe the EIT effect, two of the states should be (meta)stable. In Fig.~\ref{fig:VLevelScheme}, we assume $\ket{1}$ to be a stable ground state, and $\ket{2}$ to be metastable, i.e., $\tgamma_2$ is small. Furthermore, a strong coherent coupling $\Omega_\mathrm{c}$ between one of the (meta)stable states and the third excited state is required, e.g., between $\ket{2}$ and $\ket{3}$ in the figure. Usually, this coupling is achieved by applying a suitable control laser field. Upon weakly probing the other transition from $\ket{1}$ to $\ket{3}$ with a coherent probe field $\Omega_\mathrm{probe}$ of detuning $\Delta$ from the transition frequency, the system then exhibits transparency around $\Delta\approx 0$, i.e., $\Omega_\mathrm{probe}$ is barely absorbed at a frequency where it would be maximally absorbed in the absence of $\Omega_c$. This strongly reduced absorption is the eponymous feature of EIT.

For the EIT effect to be prominent in the presence of non-zero decay $\tgamma_2$, the inequalities 
\begin{equation}
\tgamma_3^2>|\Omega_\mathrm{c}|^2>\tgamma_3\tgamma_2
\label{eq:EITcondition}
\end{equation} 
have to be fulfilled~\cite{fleischhauer_electromagnetically_2005,rohlsberger_electromagnetically_2012,anisimov_objectively_2011}. In this equation, the first inequality can be understood as follows. The coupling field $\Omega_\mathrm{c}$ induces a splitting of the excited state into two dressed states~\cite{fleischhauer_electromagnetically_2005}. This splitting must remain smaller than the excited state linewidth $\tgamma_3$, in order to ensure that the transitions involving the two dressed states remain indistinguishable. Only then, the Fano interference characteristic of EIT appears, whereas for larger coupling fields, EIT gradually transforms into an Autler-Townes splitting without Fano interference~\cite{PhysRevLett.107.163604}. The second inequality in Eq.~(\ref{eq:EITcondition}) is found when evaluating the EIT susceptibility on resonance, and takes into account that the EIT transparency is not perfect in the presence of loss rates $\tgamma_2$. However, with the second condition fulfilled, still, a considerable reduction of the absorption is achieved. 

In order to relate the inequalities in Eq.~(\ref{eq:EITcondition}) to our discussion   of the artificial multi-level system realized with nuclei in a cavity (see Fig.~\ref{fig:ThreeLevelNuclear}), we identify the effect of the coherent coupling $\Delta_{12}$ between the excited states in the nuclear system with the externally applied coherent coupling $\Omega_\mathrm{c}$ in Fig.~\ref{fig:VLevelScheme}. Note that the incoherent coupling $\gamma_{12}$ present in the nuclear system does not appear in the archetypal EIT setup, and correspondingly should be kept negligible in the design process. Further, to connect to the discussion in Sec.~\ref{sec:three}, we rewrite Eq.~\eqref{eq:EITcondition} as
\begin{equation}
\frac{\tgamma_3^2}{\tgamma_2^2}>\frac{|\Omega_\mathrm{c}|^2}{\tgamma_2^2}>\frac{\tgamma_3}{\tgamma_2}\,.
\label{eq:EITcondSimplified}
\end{equation} 
Using this form, the observables we considered in Fig.~\ref{fig:DoubleCheese_couplings_gamma_scaledGamma_restrictedCladding} directly correspond to the quantities in the inequality. At the same time, Fig.~\ref{fig:DoubleCheese_couplings_gamma_scaledGamma_restrictedCladding} provides the means to ensure small incoherent couplings $\gamma_{12}$. Note that we will additionally consider the magnitude of the Rabi frequencies for the cavities which we select from Fig.~\ref{fig:DoubleCheese_couplings_gamma_scaledGamma_restrictedCladding}. Thus, on the basis of the level-scheme parameter ratios of Sec.~\ref{sec:invth:relative}, we will in the following choose cavities in accordance with the above conditions to design EIT dynamics in the nuclear cavity setup.

\subsubsection{Designing EIT in the artificial three-level scheme}

In order to design an EIT level scheme based on the artificial multi-level system provided by the \textit{ab initio} approach rather than on the characteristics of the observable spectrum, we have to tune this  three-level system to comply with the EIT conditions, Eq.~\eqref{eq:EITcondSimplified}. Since in principle both excited states in the artificial scheme Fig.~\ref{fig:ThreeLevelNuclear} can take equivalent roles, we have to decide which of the excited states $\ket{\mathrm{E_{1}}}, \ket{\mathrm{E_{2}}}$ should take the role of the metastable state $\ket{2}$ in Fig.~\ref{fig:VLevelScheme}.
In the following, we first consider the first excited state $\ket{\mathrm{E_1}}$, and, subsequently, the second excited state  $\ket{\mathrm{E_2}}$, as the metastable state $\ket{2}$.

With the design goal of rendering $\ket{\mathrm{E}_1}$ metastable, we for the moment adopt the notational convention 
\begin{equation}
\label{eq:1metastable}
\tgamma_2 \equiv \Gamma_\mathrm{SR, 1}+\gamma_0\,,\quad\mathrm{and}\quad\tgamma_3\equiv\Gamma_\mathrm{SR, 2}+\gamma_0\,,
\end{equation} 
according to Fig.~\ref{fig:VLevelScheme} and the general treatment of the previous Sec.~\ref{sec:genericEIT}. To fulfill the EIT requirement of Eq.~\eqref{eq:EITcondSimplified} in an optimized way, it is reasonable to consider the surface points of Fig.~\ref{fig:DoubleCheese_couplings_gamma_scaledGamma_restrictedCladding}(a). On the basis of this OS, we can firstly restrict the further analysis to the part of the surface where the scaled incoherent coupling $\gamma_{12}/(\Gamma_{\mathrm{SR,1}}+\gamma_0)$ is small. Secondly, within the range of small incoherent couplings, we can select a cavity with a decay rate ratio, $\tgamma_3/\tgamma_2=(\Gamma_{\mathrm{SR,2}}+\gamma_0)/(\Gamma_{\mathrm{SR,1}}+\gamma_0)$, and a scaled coherent  coupling strength, $\Omega_\mathrm{c}/\tgamma_2=\Delta_{12}/(\Gamma_{\mathrm{SR,1}}+\gamma_0)$, such that it complies with the EIT condition, Eq.~\eqref{eq:EITcondSimplified}. This is possible because we can essentially change the ratio of these two quantities continuously when moving on the surface of the OS in Fig.~\ref{fig:DoubleCheese_couplings_gamma_scaledGamma_restrictedCladding}(a).

Indeed, among the surface points calculated in Fig.~\ref{fig:DoubleCheese_couplings_gamma_scaledGamma_restrictedCladding}(a) a cavity that fulfils Eq.~\eqref{eq:EITcondSimplified} is readily found and indicated by a red dot. The left part of Table~\ref{tab:EIT} summarizes the properties of the corresponding level scheme. Further, relevant ratios of the properties in the convention for $\tgamma_2$ and $\tgamma_3$ adopted in Eq.~\eqref{eq:1metastable} are given in the right part of the table. Using these ratios, it is apparent that the EIT condition, Eq.~\eqref{eq:EITcondSimplified}, is well-satisfied. Furthermore, the intensity of the probe field directly driving the metastable first excited state on transition $\ket{1}\leftrightarrow\ket{2}$ in Fig.~\ref{fig:VLevelScheme} is suppressed by a factor of more than 200 as compared to the coupling on the probe transition $\ket{1}\leftrightarrow\ket{3}$ (see Table~\ref{tab:EIT}). Thus, the probing field essentially only couples to the probe transition, and the generic level scheme of Fig.~\ref{fig:VLevelScheme} is indeed realized within the nuclear platform to a good approximation. 

\begin{figure*}
\includegraphics[]{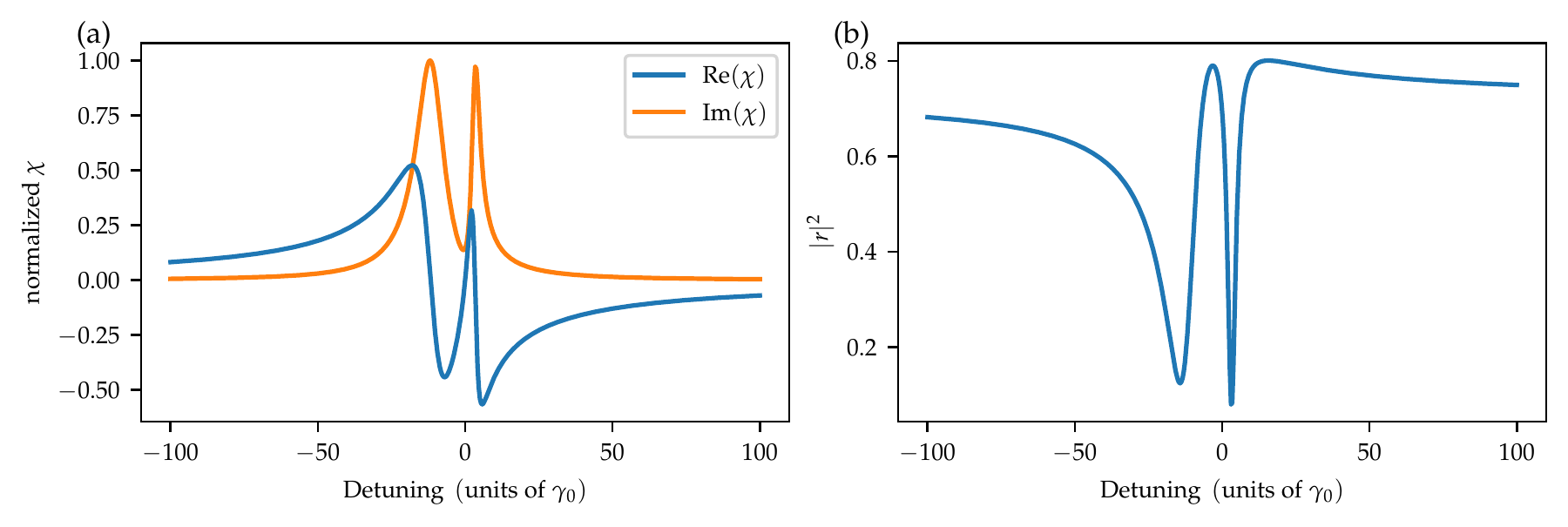}
\caption{Susceptibility and reflection spectrum for the cavity of Tab.~\ref{tab:EIT}, obtained via the inverse design of nuclear EIT setups. (a) Relevant susceptibility of the artificial level scheme probe transition (see main text for details). The susceptibility is normalized to its maximal imaginary part. The real and imaginary part show the archetypal EIT behavior. (b) Reflection spectrum as a function of the detuning from the bare nuclear transition frequency $\omega_\mathrm{nuc}$. The corresponding cavity is indicated in red in Fig.~\ref{fig:DoubleCheese_couplings_gamma_scaledGamma_restrictedCladding}(a). The double dip structure of the susceptibility imprints itself also on the reflection spectrum for the present cavity. However, the susceptibility, and not the reflection spectrum, serves as a benchmark for the successful inverse design. \label{fig:DoubleCheese_EITvsNoEIT_}}
\end{figure*}

As a witness for the successful inverse design we employ the linear susceptibility of the relevant transition between the collective nuclear states $\ket{\mathrm{G}}$ and $\ket{\mathrm{E}_2}$, which is proportional to the expectation value of the corresponding transition operator normalized to the probing field,
\begin{equation}
    \chi \propto \frac{\sigma_2^{-}(\omega)}{E_\mathrm{in}(z_2, \vec{k}_\parallel, \omega_\mathrm{nuc})}\propto -i(\vec{\mathcal{M}}^{-1})_{22}\,,\label{eq:susceptibility}
\end{equation}
using the definitions of Sec.~\ref{sec:theo} for the frequency dependent coupling matrix $\vec{\mathcal{M}}$, and neglecting the highly suppressed coupling of the probe field to the first resonant layer. For the cavity obtained above, the real and imaginary parts of the susceptibility are shown in Fig.~\ref{fig:DoubleCheese_EITvsNoEIT_}(a). Indeed, the imaginary part exhibits the archetypal double-resonance structure framing the point of (almost) transparency. Likewise, the real part shows the typical behavior with steep positive dispersion in the vicinity of the transparency window. This signature confirms that the inverse design indeed yields a three-level artificial level scheme featuring EIT-like dynamics between three states inside the cavity. 

\begin{table}
\begin{tabular}{c@{\hskip 3mm}c@{\hskip 3mm}c@{\hskip 3mm}c|c@{\hskip 3mm}c@{\hskip 3mm}c}
\hline
\hline
$\dfrac{\Gamma_{\mathrm{SR,1}}}{2}$&$\dfrac{\Gamma_\mathrm{SR, 2}}{2}$&$\Delta_{12}+i\dfrac{\gamma_{12}}{2}$&$\left|\dfrac{\tilde{\Omega}_2}{\tilde{\Omega}_1}\right|^2$&$\dfrac{\tgamma_3^2}{\tgamma_2^2}$&$\dfrac{|\Omega_\mathrm{c}|^2}{\tgamma_2^2}$&$\dfrac{\tgamma_3}{\tgamma_2}$\\[2ex]
\hline
$0.38\gamma_0$&$6.2\gamma_0$&$(6.4+0.52i)\gamma_0$&279&$58$&$13$&$7.6$
\\
\hline\hline
%\hline
\end{tabular}

\caption{Level scheme properties corresponding to the Pd($1.5\,\mathrm{nm}$)/\ch{B4C}($49.8\,\mathrm{nm}$)/\Fe{}($0.57\,\mathrm{nm}$)/\ch{B4C}($97.1\,\mathrm{nm}$)/\allowbreak\Fe{}($0.57\,\mathrm{nm}$)/\allowbreak\ch{B4C}($35.4\,\mathrm{nm}$)/\allowbreak Pd($43.7\,\mathrm{nm}$)/\allowbreak Si cavity illuminated at $2.28\,\mathrm{mrad}$. The corresponding cavity is indicated by a  red dot in Fig.~\ref{fig:DoubleCheese_couplings_gamma_scaledGamma_restrictedCladding}(a), and the linear susceptibility and the reflection spectrum are shown in Figs.~\ref{fig:DoubleCheese_EITvsNoEIT_}(a) and \ref{fig:DoubleCheese_EITvsNoEIT_}(b), respectively. The left part of the table shows the quantities determining the effective three-level scheme. $|\tilde{\Omega}_2/\tilde{\Omega}_1|^2$ is the ratio of the Rabi frequencies driving the second and first excited state, i.e., here, only the second excited state is probed. The right part of the table contains the relevant ratios according to Eq.~\eqref{eq:1metastable} that allow one to verify that Eq.~\eqref{eq:EITcondSimplified} is indeed fulfilled. We further set $\Omega_c=\Delta_{12}$, however, noting that also the inclusion of the incoherent coupling does not lead to a violation of the EIT inequality due to its comparatively small size.\label{tab:EIT}}
\end{table}

We again emphasize that the present analysis based on the \textit{ab initio} approach designs the artificial multi-level system realized inside the cavity. This approach differs from the previous approaches, which associated a multi-level scheme to the visible reflection spectrum~\cite{rohlsberger_electromagnetically_2012,heeg_collective_2015,kong_greens-function_2020,lentrodt_ab_2020}, since the observable spectra are subject to modifications by the in- and outcoupling of the probing fields. As a result, a successful realization of EIT in the multi-level system inside the cavity does not necessarily result in EIT-like signatures in the observable spectra. In turn, EIT-like signatures in the spectra do not in general imply EIT-like dynamics of the nuclei inside the cavity. In order to explore this aspect, Fig.~\ref{fig:DoubleCheese_EITvsNoEIT_}(b) shows the resulting observable reflection spectrum for this cavity as a function of the detuning from the bare nuclear transition frequency $\omega_\mathrm{nuc}$. Since the cavity is not critically coupled to the impinging x-rays, the spectrum additionally comprises a contribution due to the electronic background. The nuclear contributions manifest themselves as dips in this background, due to the interference of the different electronic and nuclear scattering channels. We stress, however, that the reflection spectrum was not of relevance for the design of the level scheme, and that it is not a necessary condition for EIT-like dynamics inside the cavity to have an EIT-like observable spectrum. 

Next, we repeat the analysis with the second possible realization of the EIT level scheme in the nuclear platform. As compared to the first approach, we exchange the roles of the states $\ket{\mathrm{E_1}}$ and $\ket{\mathrm{E_2}}$, that is we design $\ket{\mathrm{E_2}}$ as the metastable state in the level scheme in Fig.~\ref{fig:VLevelScheme}. Consequently, we now adopt the notational convention
\begin{equation}
\label{eq:2metastable}
\tgamma_2 \equiv \Gamma_\mathrm{SR, 2}+\gamma_0\,,\quad\mathrm{and}\quad\tgamma_3\equiv\Gamma_\mathrm{SR, 1}+\gamma_0\,.
\end{equation}
Note that here $\tgamma_2$ and $\tgamma_3$ are linked to the respective nuclear decay rates in a way opposite to Eq.~\eqref{eq:1metastable}, because the roles of the excited states of Fig.~\ref{fig:ThreeLevelNuclear} in the EIT setup in Fig.~\ref{fig:VLevelScheme} have changed. With the roles of the excited states exchanged, we must now consider the OS in Fig.~\ref{fig:DoubleCheese_couplings_gamma_scaledGamma_restrictedCladding}(b). On that basis, we proceed analogously to the previous part and eventually select the cavity marked by a red dot in Fig.~\ref{fig:DoubleCheese_couplings_gamma_scaledGamma_restrictedCladding}(b). The corresponding level scheme properties are given in Tab.~\ref{tab:NoEIT}, analogous to Tab.~\ref{tab:EIT} but following the convention of Eq.~\eqref{eq:2metastable}.

We find that this setting also satisfies the EIT conditions Eq.~\eqref{eq:EITcondSimplified}, though not as well as the cavity in  Tab.~\ref{tab:EIT}. However, in this configuration, the externally applied probe field does not only couple to the probe transition $\ket{1}\leftrightarrow\ket{3}$, but also to the transition $\ket{1}\leftrightarrow\ket{2}$ between the two metastable states in the EIT level scheme in Fig.~\ref{fig:VLevelScheme}, as the field intensity at the layer corresponding to the state $\ket{2}$ is suppressed only by a factor of about four in the intensity. As a result,  the implicit assumption that the probing field in EIT only couples to transition $\ket{1}\leftrightarrow\ket{3}$ in Fig.~\ref{fig:VLevelScheme} is violated. Consequently, we cannot expect the system to exhibit EIT. This undesired outcome is possible, since we did not include conditions on the Rabi frequencies into our design rules. This observation exemplifies the significance of choosing an appropriate set of design goals in order to achieve the desired functionality.

\begin{table}
\begin{tabular}{c@{\hskip 3mm}c@{\hskip 3mm}c@{\hskip 3mm}c@{\hskip 3mm}|@{\hskip 3mm}c@{\hskip 3mm}c@{\hskip 3mm}c}
\hline
\hline
$\dfrac{\Gamma_{\mathrm{SR,1}}}{2}$&$\dfrac{\Gamma_\mathrm{SR, 2}}{2}$&$\Delta_{12}+i\dfrac{\gamma_{12}}{2}$&$\left|\dfrac{\tilde{\Omega}_1}{\tilde{\Omega}_2}\right|^2$&$\dfrac{\tgamma_3^2}{\tgamma_2^2}$&$\dfrac{|\Omega_\mathrm{c}|^2}{\tgamma_2^2}$&$\dfrac{\tgamma_3}{\tgamma_2}$\\[2ex]
\hline
$3.2\gamma_0$&$0.63\gamma_0$&$(6.2+1.1i)\gamma_0$&4.4&$10$&$7.6$&$3.2$
\\
\hline\hline
\end{tabular}

\caption{Level scheme properties analogous to Tab.~\ref{tab:EIT}, but for the Pd($3.0\,\mathrm{nm}$)/\allowbreak\ch{B4C}($42.5\,\mathrm{nm}$)/\allowbreak\Fe{}($0.57\,\mathrm{nm}$)/\allowbreak\ch{B4C}($143.4\,\mathrm{nm}$)/\allowbreak\Fe{}($0.57\,\mathrm{nm}$)/\allowbreak\ch{B4C}($72.9\,\mathrm{nm}$)/\allowbreak Pd($43.4\,\mathrm{nm}$)/\allowbreak Si cavity illuminated at an angle of $2.23\,\mathrm{mrad}$. This cavity is indicated by the  red dot in Fig.~\ref{fig:DoubleCheese_couplings_gamma_scaledGamma_restrictedCladding}(b). In this setup, the probing x-rays mainly couple to the first excited state, but the coupling to the other excited state is only weakly suppressed. The conventions for the right part of the table are according to Eq.~\eqref{eq:2metastable}. We see that Eq.~\eqref{eq:EITcondSimplified} is indeed fulfilled, even though not as pronounced as for Tab.~\ref{tab:EIT}. \label{tab:NoEIT}}
\end{table}

This section showed that when design goals are chosen suitably, the inverse approach can be used to realize and optimize quantum optical multi-level systems at x-ray energies with particular desired functionalities.
For other applications, a direct design of the observable spectra may be more favorable. Therefore, in the following, we discuss the direct inverse design of the observable spectra.

\section{Inverse design of observable spectra\label{sec:reflection}}

In the following, we discuss the direct engineering of observable spectra, without the intermediate step of considering the underlying multi-level system. To this end, we combine the \textit{ab initio} Green's function approach with the diagonal representation of the spectra introduced in Sec.~\ref{sec:normalTheory}. 

For simplicity, we again focus on the minimum relevant example of cavities comprising two resonant layers. Correspondingly, the spectra consist of an electronic background which is constant in frequency as well as two Lorentzian contributions from the nuclei. First, we map out OSs for the two Lorentzian contributions to the spectrum, and show how this information can be utilized to design the spectra. As an example, we consider the realization of spectra resembling those of two-level systems -- usually associated to cavities with a single resonant layer. Subsequently, we illustrate more general designs, by tailoring the splitting between the two resonances and their relative amplitudes.

\subsection{Description of two resonant layer systems}
For the case of two resonant layers, the spectrum comprises two Lorentzian contributions. Correspondingly, the associated matrices and vectors are two-dimensional and there are two eigenvectors (also referred to as eigenmodes) of the complex level shift matrix
\begin{align}
    \vec{\mathcal{E}} &= 
    \left(
    \begin{matrix}
    \mathcal{E}_1&\mathcal{E}_{12}\\
    \mathcal{E}_{12}&\mathcal{E}_2\\
    \end{matrix}
    \right)\,,\\
    \mathcal{E}_{j}& = -\Delta_{\mathrm{CLS},j} +i\frac{\Gamma_{\mathrm{SR},j}}{2}\,, \qquad %(j\in\{1,2\})\,,\\[2ex] 
 \mathcal{E}_{12} = \Delta_{12} +i\frac{\gamma_{12}}{2}\,,
\end{align}
where we use the notation of Sec.~\ref{sec:invth}. Note that this matrix is not Hermitian, such that it may be non-diagonalizable in general, corresponding to the occurrence of an exceptional point~\cite{el-ganainy_non-hermitian_2018, miri_exceptional_2019}. In the following, we assume that the matrix can be diagonalized. The eigenvalues evaluate to
%\begin{align}
%\lambda_{1,2} = &-\frac{ \Delta_{\mathrm{CLS,1}}+\Delta_{\mathrm{CLS,2}} }{2}+i\frac{\Gamma_{\mathrm{SR,1}}+\Gamma_{\mathrm{SR,2}}}{4} \nonumber \\
%%
%&\mp\frac{1}{2} \left[\left(\Delta_{\mathrm{CLS,2}}-\Delta_{\mathrm{CLS,1}}+i\frac{\Gamma_{\mathrm{SR,1}}-\Gamma_{\mathrm{SR,2}}}{2}\right)^2 \right. \nonumber \\
%%
% & \qquad \qquad \left . +4\left(\Delta_{12}+i\frac{\Gamma_{12}}{2}\right)^2 \right]^{\frac 12}\,, 
%\label{eq:invth:lambda12}
%\end{align}
%
\begin{align}
\lambda_{1,2} &= \frac{ \mathcal{E}_{1}+\mathcal{E}_{2} }{2} \mp\frac{1}{2} \sqrt{\left(\mathcal{E}_{1}-\mathcal{E}_{2}\right)^2  +4\,\mathcal{E}_{12}^2  }\,. \label{eq:invth:lambda12}
\end{align}
It is important to note that upon varying the cavity structure or the incidence angle, the association between the eigenvalues and their labels $1,2$ is not necessarily preserved. The different rates entering the level shift matrix $\mathcal{E}_{ll'}$ are continuous functions of the cavity parameters. However, whenever the argument of the square root function in Eq.~(\ref{eq:invth:lambda12}) crosses the branch cut, the labeling is interchanged, such that artificial discontinuities arise in the respective eigenvalues as a function of the cavity parameters.

\begin{figure*}
\includegraphics[]{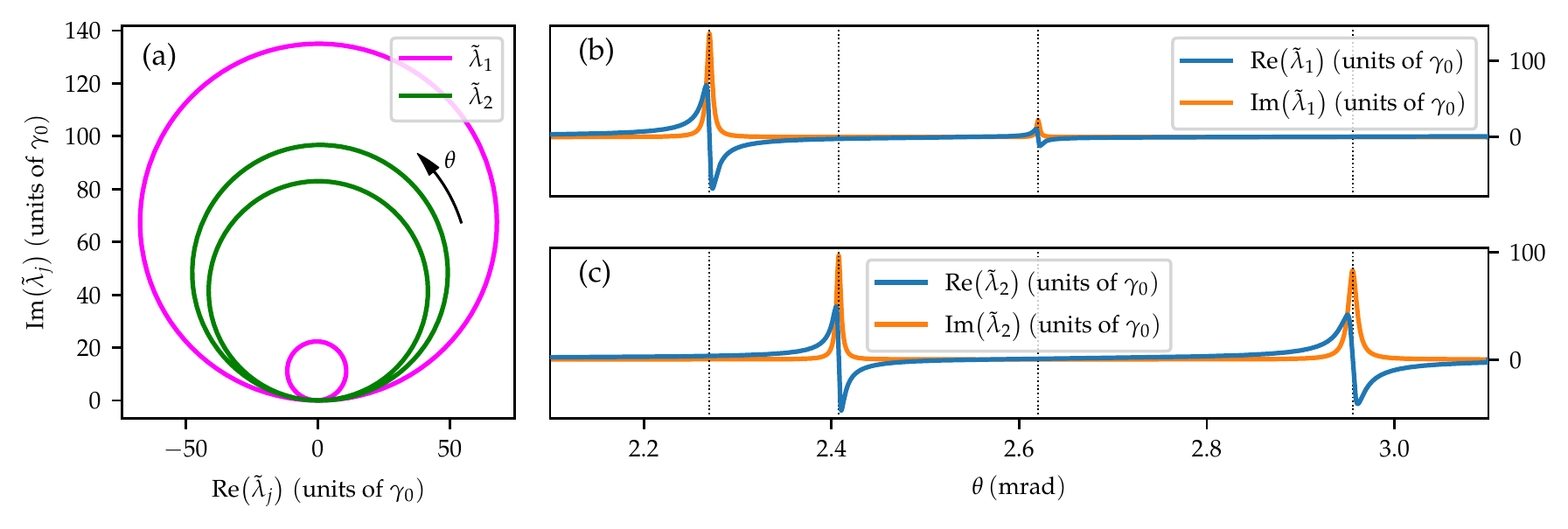}
\caption{Eigenmode structure as a function of the incidence angle. (a) Trajectories of the diagonal basis eigenvalues in the complex plane as a function of the angle of incidence for the cavity of Fig.~\ref{fig:PoleStructureCouplings}. The arrow indicates the direction of increasing angle. (b,c) Explicit dependency of the real and imaginary contribution to the eigenvalue on the angle of incidence. Note that discontinuities arising from the ambiguity in the labeling $1,2$ of the eigenvalues upon passing through a branch cut in Eq.~\eqref{eq:invth:lambda12} are removed in this figure. For this, $\lambda_1$ and $\lambda_2$ are exchanged at each crossing of the branch cut, indicated by dashed lines in (b,c). \label{fig:DoubleCheese_diagonalizedTrajectories}}
\end{figure*}

In order to illustrate the eigenvalue structure,  Fig.~\ref{fig:DoubleCheese_diagonalizedTrajectories} shows the  eigenvalues as a function of the angle of incidence for the cavity of Fig.~\ref{fig:PoleStructureCouplings} as an example. 
In this figure, we use alternative quantities $\tilde{\lambda}_1(\theta), \tilde{\lambda}_2(\theta)\in\lbrace\lambda_1(\theta), \lambda_2(\theta)\rbrace$ for which the discontinuities in the eigenvalues are removed by suitably interchanging $\lambda_1(\theta) \leftrightarrow \lambda_2(\theta)$ upon each crossing of the branch cut.
These crossings are indicated by the dotted lines in panels (b,c). 
The trajectories of the eigenvalues can be interpreted in analogy to the CLS and SR in Figs.~\ref{fig:ThreeLevelSysBareCouplings} and \ref{fig:PoleStructureCouplings}. However, one of the eigenvectors in the diagonal basis only couples to the odd modes of the cavity, while the second one solely couples to even modes. Therefore, each eigenvalue only has resonances for every second Green's function pole, see Fig.~\ref{fig:DoubleCheese_diagonalizedTrajectories}(b,c). As a result, the resonances are further apart, thereby improving the single mode approximation of Eq.~\eqref{eq:ML_SM}. This explains the circular appearance of the trajectories in Fig.~\ref{fig:DoubleCheese_diagonalizedTrajectories}(a).

%Hence, for the following discussion, we have to keep in mind that the indices of $\lambda_{1,2}$ do not uniquely relate to the first and second Lorentzian contribution to the spectrum, but may change their assignment for different cavity geometries and angles of incidence. Yet, this is merely a technical aspect that does not affect the interpretation of the eigenvalue trajectories in Fig.~\ref{fig:DoubleCheese_diagonalizedTrajectories}. 

%In fact, finding the exceptional points turns out to be rather straightforward as the spectral weights diverge for that case, $|g_n|\rightarrow\infty$. The resulting spectrum is still meaningful as the spectral weights asymptotically behave like $g_1\sim-g_2$ and hence the divergence cancels for coinciding eigenvalues $\lambda_1=\lambda_2$. Exceptional points are hence found upon maximizing the absolute value of one of the spectral weights.

\begin{figure}%
\includegraphics[]{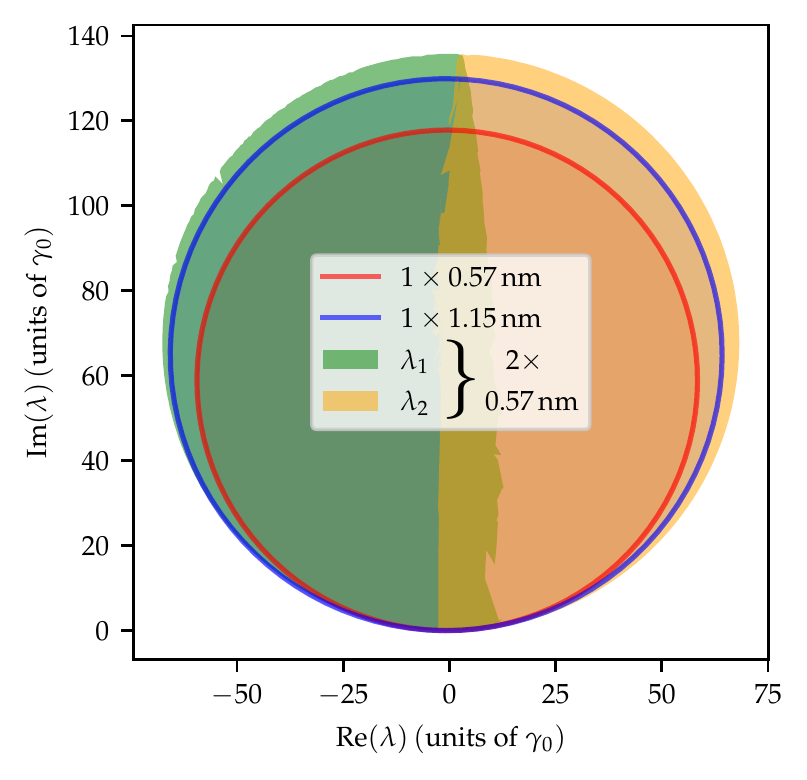}%
\caption{Accessible OS for the eigenvalues in the diagonal representation of the spectrum for a Pd/\allowbreak\ch{B4C}/\allowbreak\Fe{}$(0.57~\mathrm{nm})$/\ch{B4C}/\allowbreak\Fe{}$(0.57~\mathrm{nm})$/\allowbreak\ch{B4C}/\allowbreak Pd/\allowbreak Si cavity setup. The green and orange shaded areas show the accessible OSs for the individual eigenvalues $\lambda_{1,2}$ according to Eq.~\eqref{eq:invth:lambda12}. Note that the separation into $\lambda_1$ and $\lambda_2$ is not unique. The red line corresponds to the extremal \CLS and \SR/2 that are accessible by a cavity with a single resonant layer within an archetype Pd/\allowbreak\ch{B4C}/\allowbreak\Fe{}$(0.57~\mathrm{nm})$/\allowbreak\ch{B4C}/\allowbreak Pd/\allowbreak Si cavity setup. Likewise, the blue line corresponds to the extremal \CLS and \SR/2 that are accessible by the single-resonant-layer cavities of twice the resonant layer thickness within an archetype Pd/\allowbreak\ch{B4C}/\allowbreak\Fe{}$(1.15~\mathrm{nm})$/\allowbreak\ch{B4C}/\allowbreak Pd/\allowbreak Si cavity setup. We see that cavities with two-resonant layers provide a larger accessible parameter space than the corresponding single-resonant-layer cavities, even if the same amount of resonant material is considered in both cases.\label{fig:doubleCheese_lambdaMP}}%
\end{figure}

As a first step towards the design of the spectra, we consider the range of eigenvalues $\lambda_{1,2}$ accessible with the Pd/\ch{B4C}/\Fe{}$(0.57~\mathrm{nm})$/\ch{B4C}/\allowbreak\Fe{}$(0.57~\mathrm{nm})$/\allowbreak\ch{B4C}/\allowbreak Pd/\allowbreak Si cavity setup. The OS for each of the two eigenvalues individually is shown in Fig.~\ref{fig:doubleCheese_lambdaMP}. Since the assignment of the eigenvalue indices 1,2 is not unique but changes at each branch cut in Eq.~\eqref{eq:invth:lambda12} as discussed above, the joint set of both eigenvalues indicates the eigenvalue tuning range. Considering the trajectories in Fig.~\ref{fig:DoubleCheese_diagonalizedTrajectories} as function of the incidence angle, is is not surprising that the total OS in Fig.~\ref{fig:doubleCheese_lambdaMP} is also found to be circular.

Interestingly, the eigenvalues of the cavity with two layers of M\"ossbauer nuclei can have imaginary parts of up to about $135\gamma_0$. Opposed to that, the corresponding Pd/\ch{B4C}/\allowbreak\Fe{}$(1.15~\mathrm{nm})$/\allowbreak\ch{B4C}/\allowbreak Pd/\allowbreak Si cavity setup in which the same amount of the M\"ossbauer nuclei is combined into a single layer has a smaller OS. The single-resonant-layer system with a ${}^{57}$Fe thickness of 0.574~nm is even further restricted in the range of accessible CLS and SR. In order to illustrate this, we plot the boundaries of the accessible OS for the single-resonant-layer systems with layer thickness 0.574~nm [1.15~nm] as red [blue] lines in Fig.~\ref{fig:doubleCheese_lambdaMP}. For this comparison, however, one has to note that thicker iron layers support long-range magnetic ordering, and the resulting hyperfine field leads to a splitting of the single resonance into the magnetic substructure. Furthermore, field gradients across thicker layers may lead to corrections to the two-level spectra considered in the present work~\cite{lentrodt_ab_2020}.

In the following, we use the eigenmode properties discussed above to design observable reflection spectra.

\subsection{Design of single-resonance spectra}
\begin{figure*}
%\begin{captionbeside}[] %
\includegraphics[]{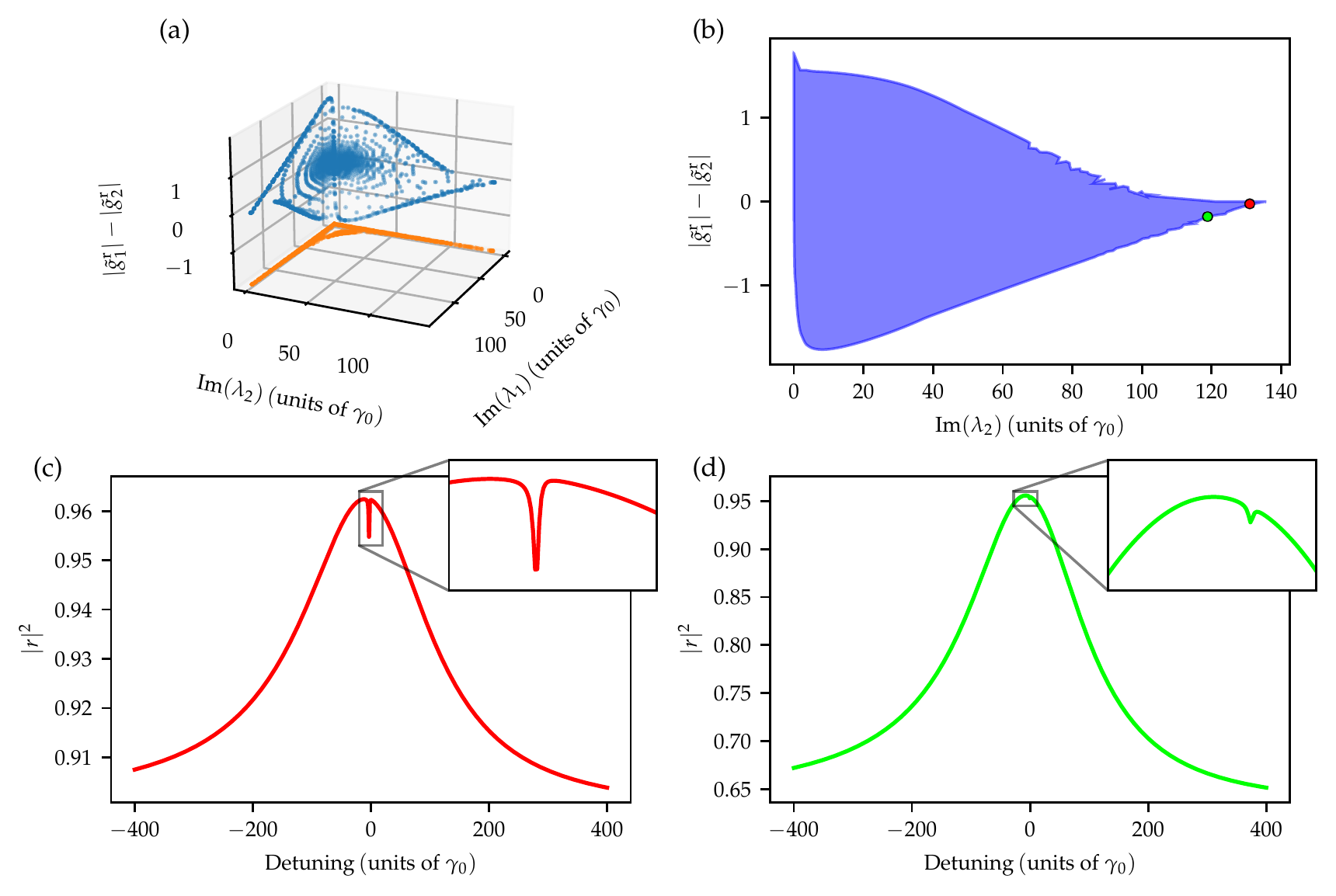}
\caption{\label{fig:DoubleCheese_ImLM_ImLP_scaledWeightDifference}Design of two-level-like spectra with cavities containing two resonant layers. (a) Accessible OS for the 
difference in the the scaled spectral weights of the two eigenmodes, plotted against their linewidth enhancements (SR). The blue dots indicate the computed surface points of the set of accessible combinations. Orange dots show the projection of the blue points onto the plane spanned by the imaginary parts of the eigenvalues. (b) Projection of the surface of (a) onto the plane normal to the $\Im(\lambda_1)$ direction. The red and green dots indicate the cavities used for panels (c) and (d), respectively. (c) Reflected intensity as a function of the detuning from the bare nuclear transition frequency for a Pd($9.3\,\mathrm{nm}$)/\ch{B4C}($20.6\,\mathrm{nm}$)/\Fe{}($0.57\,\mathrm{nm}$)/\ch{B4C}($26.2\,\mathrm{nm}$)/\allowbreak\Fe{}($0.57\,\mathrm{nm}$)/\allowbreak\ch{B4C}($20.6\,\mathrm{nm}$)/\allowbreak Pd($17.0\,\mathrm{nm}$)/\allowbreak Si cavity setup probed at an angle of $2.29\,\mathrm{mrad}$. The dominant eigenvalue $\lambda_2=(3.2+130.6i)\gamma_0$ shows a linewidth enhancement larger than what is achievable with only one resonant layer in an analogous setting. (d) Analogous to panel (c), but for an Pd($6.5\,\mathrm{nm}$)/\ch{B4C}($27.0\,\mathrm{nm}$)/\Fe{}($0.57\,\mathrm{nm}$)/\ch{B4C}($2.6\,\mathrm{nm}$)/\allowbreak\Fe{}($0.57\,\mathrm{nm}$)/\allowbreak\ch{B4C}($31.2\,\mathrm{nm}$)/\allowbreak Pd($16.7\,\mathrm{nm}$)/\allowbreak Si cavity  probed at an angle of $2.32\,\mathrm{mrad}$, where the relevant eigenvalue is $\lambda_2=(0.55+118.8i)\gamma_0$. The linewidth enhancement is very close to the ones achievable with a single resonant layer in an analogous setting, but the visibility is greatly enhanced.}%
%\end{captionbeside}
\end{figure*}

From the examples in Secs.~\ref{sec:micromacro}, \ref{sec:analysisSymmetric} and \ref{sec:invth:EIT}, we learned that under suitable conditions, the imprint of one of the eigenmode contributions to the reflection spectrum can be highly suppressed. While this behavior may be undesired for most cases, we may also use it to design a two-level system-like response with only one resonance in the spectrum, using a system comprising several resonant layers. The motivation for this is that Fig.~\ref{fig:doubleCheese_lambdaMP} showed that the eigenvalues characterizing the Lorentzian contributions to the spectrum allow for higher SRs and CLSs than what is feasible with a corresponding single resonant layer system. We can therefore envision the design of a single line spectrum when one Lorentzian is strongly suppressed while the other shows properties beyond the single resonant layer capabilities.  Indeed, in the following, we show that the design of two-level-system-like reflection spectra with SR exceeding the tuning range of corresponding single-resonant-layer systems is possible. 

To that end, we consider the weights $g^{\mathrm{r}}_n$ of both Lorentzians in the reflection spectrum [see Eq.~\eqref{eq:reflection}]. In order to evaluate the visible impact of each eigenmode onto the spectrum, we scale the weight by the corresponding linewidth, \begin{equation}
\tilde{g}^{\mathrm{r}}_n=\frac{g^{\mathrm{r}}_n}{\mathrm{Im}(\lambda_n)+\gamma_0/2}.
\end{equation} 
To suppress one eigenmode while having maximum weight for the other one, we consider the difference of the corresponding weights ($|\tilde{g}^{\mathrm{r}}_1|-|\tilde{g}^{\mathrm{r}}_2|$) as a design goal. When this quantity is positive and large, mainly $\lambda_1$ contributes to the spectrum. Likewise, the influence of $\lambda_1$ is suppressed in comparison to $\lambda_2$ for large and negative values. Furthermore, the difference has the crucial property that it does not diverge when the complex level shift matrix $(\mathcal{E}_{ll'})=(\Delta_{ll'}+i\gamma_{ll'}/2)$ becomes non-diagonalizable. Reaching such an exceptional point, the eigenvalues become identical $\lambda_1\rightarrow\lambda_2$, while the spectral weights may acquire opposite phases and asymptotically tend to the same large absolute value, $g^{\mathrm{r}}_1\sim-g^{\mathrm{r}}_2$.  

For the design, we augment the scaled weight difference design goal with imaginary parts of both eigenvalues, corresponding to the spectral linewidths of the eigenmodes. This allows us to design the magnitude of the contribution to the spectrum together with the linewidth enhancements. 

The surface of the accessible OS for the above parameters is shown in Fig.~\ref{fig:DoubleCheese_ImLM_ImLP_scaledWeightDifference}(a) as the blue ``spiderweb-like'' structure. Note that even though the assignment of the eigenvalue indices $1,2$ is not unique as discussed above, our sampling is complete as long as we consider both eigenvalues in a single OS. We show the calculated surface points for the OS in the figure, rather than combining them into a closed surface, in order to better illustrate the structure of the OS. From the projection of the OS onto the plane of the imaginary parts (shown in orange), it is clear that when the linewidth of one Lorentzian is large, the other one necessarily tends to be small. The OS thus has a bimodal structure in which either $\lambda_1$ or $\lambda_2$ has the large imaginary part. For our optimization, we choose $\lambda_2$ to take the dominant role and suppress the influence of the remaining eigenvalue. As the imaginary part of the suppressed eigenvalue is not of significance for the following discussion, we project the OS onto the plane normal to $\Im(\lambda_1)$ in Fig.~\ref{fig:DoubleCheese_ImLM_ImLP_scaledWeightDifference}(b). From this figure, we find that upon approaching large linewidths, the tuning capabilities for the scaled weight difference substantially decrease and eventually collapse to differences close to zero. 

For the interpretation, we consider the  reflection spectra of two cavities in the following, which are marked  in Fig.~\ref{fig:DoubleCheese_ImLM_ImLP_scaledWeightDifference}(b) by a red and green dot, respectively. Figure~\ref{fig:DoubleCheese_ImLM_ImLP_scaledWeightDifference}(c) shows the spectrum corresponding to the cavity marked by a red dot. Its main resonance has a linewidth of about $131\,\gamma_0$. The influence of the second eigenmode is still visible as a small dip close to zero detuning, but is largely suppressed as compared to the spectral imprint of the broad Lorentzian. Notably, the linewidth realized is not only larger than what is achievable with a single resonant layer of the same thickness ($0.574~\mathrm{nm}$) but indeed exceeds the ones achievable with a resonant layer thickness of $1.15~\mathrm{nm}$ (see Fig.~\ref{fig:doubleCheese_lambdaMP}). Thus, larger tunability is available with less restrictions regarding intrinsic magnetic ordering and field inhomogeneities within the resonant layer.

Likewise, Fig.~\ref{fig:DoubleCheese_ImLM_ImLP_scaledWeightDifference}(d) shows the spectrum of the  cavity marked with a green dot in (b). It's main resonance has a linewidth enhancement of $119\,\gamma_0$, which is very close to the maximally achievable value for cavities with a single thin resonant layer~\cite{diekmann_inverse_2022}. As compared to the cavity in (c), this one has a slightly smaller linewidth enhancement, but at the same time allows for a better tuning of the eigenmode weight difference, see panel (b). As a result,  the second eigenmode can be suppressed by orders of magnitude. 
Note that in principle, a similar SR would be accessible with the archetypal systems comprising a single resonant layer, which were considered in~\cite{diekmann_inverse_2022}. However, for single-resonant-layer systems with SR close to the maximally achievable one, the visibility of the nuclear signature in the reflection spectrum is strongly suppressed. As a result, the largest SRs are experimentally not accessible in single-resonant-layer systems.
In contrast, the two-resonant-layer system in Fig.~\ref{fig:DoubleCheese_ImLM_ImLP_scaledWeightDifference}(d) exhibits a clear and experimentally accessible signature of the nuclear response in the reflection spectrum. 

In Fig.~\ref{fig:DoubleCheese_diagonalizedTrajectories}, we found that cavities featuring a large superradiant enhancement also exhibit correspondingly large collective line shifts, at least when the single mode approximation \eqref{eq:ML_SM} is applicable~\cite{diekmann_inverse_2022}. The CLS and SR then transition into each other as a function of the x-ray angle of incidence. Also for the cavities with large linewidth enhancement presented in Fig.~\ref{fig:DoubleCheese_ImLM_ImLP_scaledWeightDifference}, correspondingly high frequency shifts can be realized by suitably adjusting the angle of incidence. In this process, the suppression of the second eigenmode contribution remains approximately constant. As a result, we find that the enhanced tuning capabilities offered by the two-resonant-layer cavities carry over to the design of collective frequency shifts.

In summary, we have shown that the spectra of cavities with two resonant layers  can be designed in such a way that they resemble those of two-level systems. Usually, such two-level spectra are associated to cavity structures with a single resonant layer only. However, the implementation of tunable two-level spectra with cavities comprising two resonant layers has significant advantages. First, the tuning range for the CLS and SR are larger. Even more importantly, we could show that there are SRs and CLSs within the enlarged tuning range in two-resonant-layer cavities which are experimentally accessible. This is not trivial, as the visibility of the nuclear signature was found to vanish for single-resonant-layer cavities with highest CLS and SR~\cite{diekmann_inverse_2022}. These qualitative advantages are enabled by the cavity-mediated coupling between the two layers.

Interestingly, in Fig.~\ref{fig:DoubleCheese_ImLM_ImLP_scaledWeightDifference}(b), the geometry of the cavities with near-maximum SR and CLS is not uniquely fixed by the design goals considered so far. In the next section, we exploit the comparably large number of cavity parameter degrees of freedom to optimize the resulting spectra even further towards particular shapes.

\subsection{Towards the design of arbitrary reflection spectra}

In the previous section, we showed that the accessible OSs for the parameters of the spectrum can be mapped out, and considered the design of two-level system-like spectra. However, this design was comparably simple in the sense that it only required the consideration of few parameters. Here, we consider more general designs of the reflection spectrum. To this end, we exploit more of the degrees of freedom determining the overall spectrum, such as the relative magnitude of the eigenmode weights and the linewidth enhancements and frequency shifts of both Lorentzians. In order to obtain a tractable optimization problem,  we define a suitable cost function that comprises contributions from all the desired features of the spectrum. The optimization of this cost function allows one to directly identify cavity structures which realize the desired spectral signatures, bounded only by the physical limitations of the platform. 

\subsubsection{Cost function}
To illustrate the approach, we design spectra with two well-separated symmetrical dips, clearly visible in a large electronic background in the reflection spectrum. The considered cost function is
\begin{align}
f(\Delta, \alpha) = &-2\Big|\left|\Re\lambda_1-\Re\lambda_2\right|-\Delta\Big|\notag\\[1ex]
&-15\left|\frac{|\tilde{g}^\mathrm{r}_1|}{\sqrt{1-\sqrt{\alpha}}}-|\tilde{g}^\mathrm{r}_2|\sqrt{1-\sqrt{\alpha}}\right|\notag\\[1ex]
&+2\left({|\tilde{g}^\mathrm{r}_1|}+|\tilde{g}^\mathrm{r}_2|\right)\left|\frac{\Re\lambda_1}{\Im\lambda_1+\gamma_0/2} \right.\notag\\[1ex]
&\left. \qquad -\frac{\Re\lambda_2}{\Im\lambda_2+\gamma_0/2}\right|\notag\\[1ex]
&-1.5\left|\Im(\lambda_1)-\Im(\lambda_2)\right|\notag\\[1ex]
&+0.05\left|r_{\mathrm{el}}\right|\,.
\label{eq:costfunctionSplitting}
\end{align}
\begin{figure*}
%\begin{captionbeside}[] %
\includegraphics[]{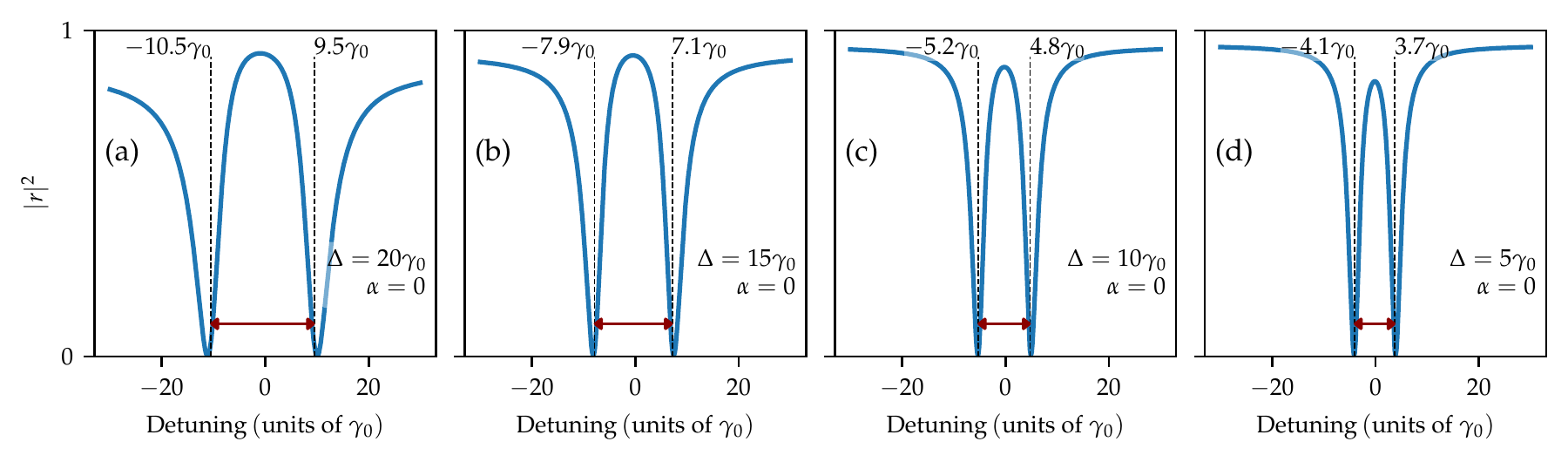}
\caption{\label{fig:Double_Cheese_Splitting} Direct design of the reflection spectra. In this example, spectra with two pronounced dips in the off-resonant electronic background are designed with a specific separation $\Delta$ between the dips. The spectra were obtained by maximizing the cost function~\eqref{eq:costfunctionSplitting} with the respective value of $\Delta$ over the cavity parameters. The central position of the two eigenmode contributions are indicated by dashed lines and their absolute position is annotated at the top of these lines. Note that the dashed lines do not precisely coincide with the minima of the dips due to interference with the tails of the neighboring spectral lines. Remarkably, the cavities associated to the above spectra all come without upper cladding layer.}%
%\end{captionbeside}
\end{figure*}
Here, the first term constrains the distance $\Delta$ of the two dips in the reflection spectrum. The second term enforces a depth $\alpha$ of the dip towards higher detunings. The third term is responsible for non-vanishing weights, i.e., it ensures that the dips are well-visible in the spectrum. In the previous section we already mentioned that a maximization of spectral weights alone is not meaningful as these weights may diverge when the matrix $(\Delta_{ll'}+i\gamma_{ll'}/2)$ is not diagonalizable. These exceptional points imply that the two eigenvalues of the matrix coincide. We make use of this by multiplying the spectral weights with the difference of the frequency shifts scaled by their respective linewidths. As a result, the divergence upon reaching such an exceptional point is suppressed. At the same time, the second part of the third term ensures a comparatively large separation on the scale of the linewidth of the Lorentzian contributions. The fourth term is used to enforce two dips of similar width  in the spectrum. Finally, the last term sets a preference for a large electronic background.

Obviously, constructing such a cost function involves the fine tuning of the different terms in order to reach the desired final spectra. We stress, however, that this fine tuning of the cost function is in terms of the properties of the observable spectrum, and not in terms of the cavity parameters. The efficient maximization of such a cost function then relies on a numerically efficient implementation of the involved quantities via the Green's function description~\cite{diekmann_inverse_2022}.

\subsubsection{Design of the frequency splitting}

To begin with, we fix $\alpha=0$ and maximize the above cost function over the cavity parameters for four different values of $\Delta$. Thus, we aim at designing spectra featuring dips of equal depths ($\alpha=0$) at  well-defined distances $\Delta$.  The resulting spectra are shown in Fig.~\ref{fig:Double_Cheese_Splitting}. The respective values for $\Delta$ are indicated in each panel. We find that overall, the optimization results in the desired spectral features. However, a closer analysis reveals that only the spectra in  Figs.~\ref{fig:Double_Cheese_Splitting}(a,b,c) feature the respective desired distance $\Delta$ quantitatively. In the spectrum in (d), the dip distance differs from the design value, but follows it qualitatively. The reason for this is that the limited cavity tuning capabilities impose a trade-off between different design goals incorporated in the cost function. For example, we found that a distance of $\Delta=5\gamma_0$ can indeed be achieved by changing the relative weighting of the different cost function contributions. However, this more precise tuning of the distance comes at the expense of, e.g., a decreased symmetry of the two dips.

\begin{figure*}
%\begin{captionbeside}[] %
\includegraphics[ ]{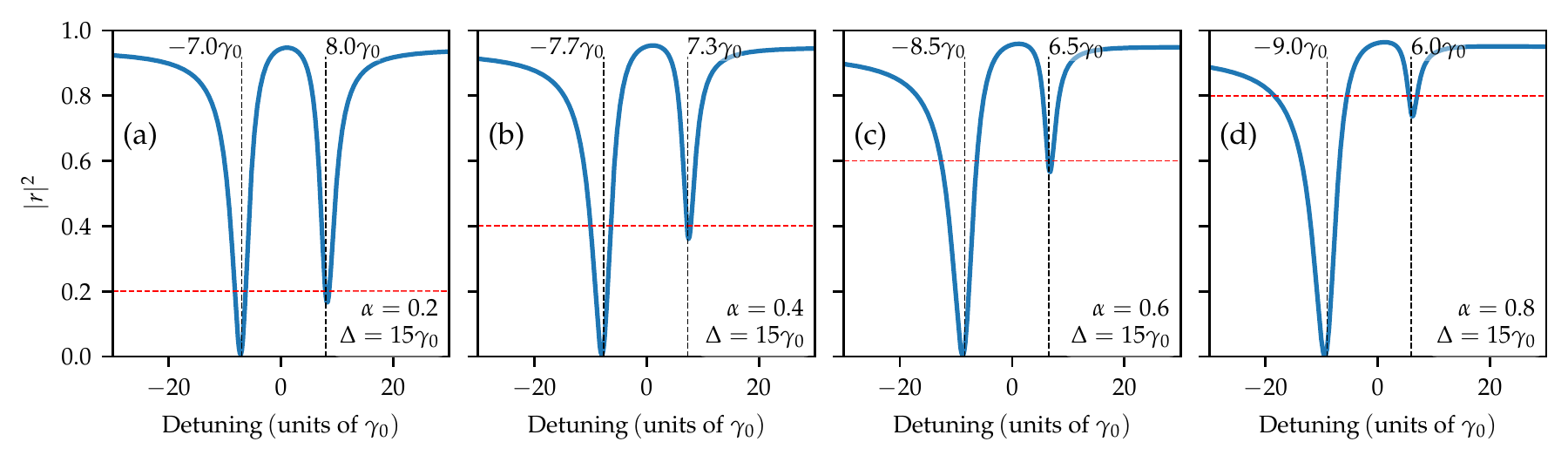}
\caption{\label{fig:Double_Cheese_Splitting_unbalanced} Design of the resonance amplitudes in the reflection spectrum. Here, the cost function is parametrized to yield two dips with a separation of $\Delta=15\gamma_0$ and varying depth $\alpha$ of the resonance at positive detunings. In contrast, the depth of the resonance at negative detunings is designed to remain constant. }%
%\end{captionbeside}
\end{figure*}

\subsubsection{Design of spectral amplitudes}
In the next step, we fix the relative distance of the two dips, and consider the design of their relative depth. In particular, we aim at changing the depth of one dip, while preserving that of the other one. This property is parametrized by the parameter $\alpha$ in the cost function \eqref{eq:costfunctionSplitting}. The particular way $\alpha$ is included in the cost function has several reasons. First, taking the square root of $\alpha$ is reasonable since we consider the amplitudes in the cost function, but the intensity in the spectra. Second, since the magnitude of the electronic background contribution is of order one, we enforce  a ratio $1-\sqrt{\alpha}$ between the scaled spectral weights in order to determine the depth of one of the dips. Last, as a technical detail, we distribute the impact of this factor onto the two modes by dividing or multiplying with $\sqrt{1-\sqrt{\alpha}}$ in order to keep the balance with the other terms in the cost function when different values of $\alpha$ are inserted. 

Results of the optimization for different values of $\alpha$ are shown in Fig.~\ref{fig:Double_Cheese_Splitting_unbalanced}. We find that, as desired, the left dip remains almost unaffected across the different spectra, while the right one is reduced in depth in accordance with the respective values of $\alpha$. Note that residual  deviations of the actual depth from the value of $\alpha$ arise because the electronic reflection coefficient is not precisely of magnitude one, and from an additive contribution due to the tail of the other resonance. Such contributions could also be included when using a more complicated cost function.

It is important to note that in all four cases, the depth of the right dip is suppressed. This can be understood by noting that for $0<\alpha<1$, the cost function \eqref{eq:costfunctionSplitting} suppresses the spectral weight corresponding to $\lambda_1$. From Fig.~\ref{fig:doubleCheese_lambdaMP}, we infer that the real part of $\lambda_1$ is mostly negative. This in particular holds true for spectral separations between the two eigenmodes as large as $\Delta=15\gamma_0$, which cannot be accomplished in the overlap region between $\lambda_1$ and $\lambda_2$. As the negative real part of $\lambda_1$ in turn corresponds to a positive frequency shift of the corresponding resonance, we find the right dip to be suppressed.

\begin{figure*}
%\begin{captionbeside}[] %
\includegraphics[]{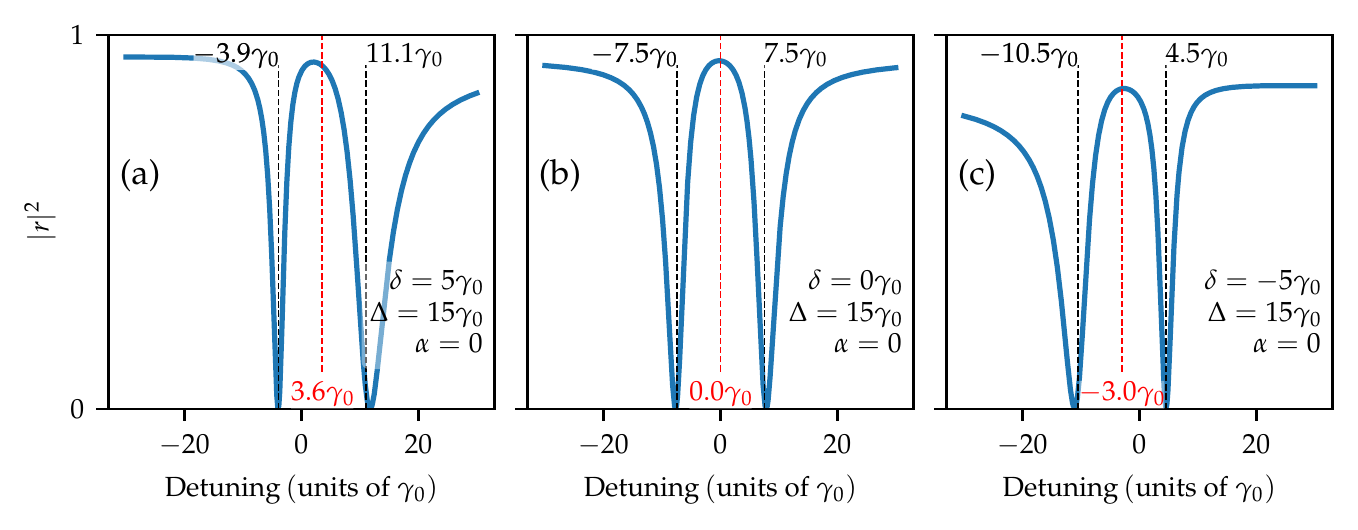}
\caption{\label{fig:Double_Cheese_Splitting_shifted} Design of absolute shifts of the reflection spectrum. Here, the cost function Eqs.~\eqref{eq:costfunctionSplitting} combined with \eqref{eq:contrCostFunction} is parametrized to yield a spectrum with two resonances separated by $\Delta=15\gamma_0$ at different absolute position $\delta$. The three $\delta$ values chosen are indicated in the respective panels. }%
%\end{captionbeside}
\end{figure*}

\subsubsection{Design of overall spectral shifts}
Finally, we extend our design towards  the absolute positioning of the double-dip structure. To that end, we add an additional term to the cost function~\eqref{eq:costfunctionSplitting},
\begin{equation}
-4\left|\frac{\Re\lambda_1+\Re\lambda_2}{2}+\delta\right|.
\label{eq:contrCostFunction}
\end{equation}
This contribution is maximized, when the center of the two eigenmode contributions is shifted by $\delta$. Note that the negative real parts of $\lambda_j$ describe the eigenmode frequency shifts.

The results of the optimization procedure are shown in Fig.~\ref{fig:Double_Cheese_Splitting_shifted} for different values of $\delta$. The other parameters related to the dip splitting and their relative amplitude are kept unchanged. We find that the position of the resulting overall spectra indeed follows the changes in $\delta$. However, the realized shifts do not precisely correspond to the respective values of $\delta$, and the shifts go along with modifications in the line shapes of the two dips in panels (a) and (c). Again, this can be attributed to the limited overall tuning capabilities of the considered cavity structure.

Summarizing, we therefore conclude that while the design of the spectrum naturally is subject to limitations imposed by the tuning capabilities of the chosen cavity structure, the inverse approach presented above allows one to efficiently navigate within these restrictions. It thus provides a flexible tool to fully exploit the design capabilities of the reflection spectrum for particular applications. Other observables such as the transmission spectrum can be designed analogously.

\section{Discussion and Outlook}\label{sec:conc}

Layers of resonant M\"ossbauer nuclei embedded in thin-film cavities form a promising platform for x-ray quantum optics. In the low-excitation regime, these systems feature a simple description in terms of artificial multi-level schemes. In previous works, it has been demonstrated that this artificial multi-level approach allows one to interpret the observable spectra of such nuclei-cavity systems in terms of well-known quantum optical phenomena. However, even though there are several established approaches to calculate observables for a given cavity structure, the inverse process of determining a cavity structure delivering a desired functionality remained an open challenge so far. As a result, it was unknown which quantum optical setups can be implemented in general using the thin-film cavity setups, and which functionalities they can provide. For cavity structures with a single layer of nuclei, giving rise to artificial two-level systems, this question was answered in~\cite{diekmann_inverse_2022} using an inverse design approach.

Motivated by this result, here, we developed the inverse design of more general thin-film cavities featuring multi-level schemes, with the goal of determining cavity structures which deliver a specified functionality in an optimal way. 

Throughout the analysis, we found that the established approach of associating a multi-level scheme to a given cavity structure based on an observable, such as the reflection spectrum, is not unique. Instead, an equivalent analysis based on another observable, such as the transmission spectrum, may lead to a different, and potentially incompatible, underlying multi-level system for the same cavity structure. Our analysis identified the coupling of the x-rays into and out of the cavity as the origin of these discrepancies. 

Therefore, we alternatively considered an interpretation in terms of a multi-level system predicted from the \textit{ab initio} analysis of the cavity setup. This interpretation is unique, since it relates to the artificial multi-level system realized inside the cavity, which is independent of the in- and outcoupling. The latter are only considered when calculating the observable spectra from the multi-level system dynamics in a second step.  

It is important to note that the different approaches agree in their predictions for all spectral observables -- it is the interpretation in terms of a multi-level system which differs.  
However, the different interpretations become relevant, e.g., if one is interested in the response of the system to parameter changes. For example, if effects of higher x-ray intensities on the interaction of the x-rays with the nuclei are to be considered, then an extrapolation of the interpretation in terms of a level scheme derived from the observable spectra towards higher intensities is likely to fail, since only the nuclear dynamics is affected by the higher x-ray intensity, but not the linear in- and outcoupling. 

Nevertheless, both approaches are of relevance, and complement each other regarding potential applications. The direct inverse design of the observable spectra is favorable, e.g., if a particular spectral or temporal response of the cavity is desired. In this case, it may not even be of relevance what the underlying multi-level system is. Examples include the optimization of filters~\cite{rohlsberger_grazing_1994}, x-ray dispersion control~\cite{heeg_tunable_2015}, or in general the shaping of x-ray pulses in particular ways. On the other hand, the inverse design of the \textit{ab initio} artificial multi-level system appears favorable if a particular quantum optical effect is to be realized. For this reason, we considered the direct inverse design of the observable spectra and the inverse design of the underlying \textit{ab initio} artificial multi-level system separately.

Regarding the \textit{ab initio} level schemes, we extended the previous inverse design approach presented in~\cite{diekmann_inverse_2022} to the case of thin-film cavities with several resonant layers. The associated artificial level schemes can form the basis of advanced quantum optical setups in the x-ray regime that are unavailable otherwise. Using the example of systems comprising two resonant layers we demonstrated that the inverse approach allows one to explore and exploit the full tuning potential of the different level scheme parameters. In particular, it allows one to engineer desired parameter ratios which determine the quantum optical functionality of the level scheme. We exemplified this using the design of level schemes featuring electromagnetically induced transparency based on given requirements for the parameter ratios. It is important to note that a successful implementation of a quantum optical level scheme may not lead to observable signatures compatible with those expected from the quantum optical treatment of atoms in free space, because of the influence of the outcoupling of the x-rays from the cavity. On a more general level, our analysis confirmed the qualitative design rule already found in the design of two-level systems~\cite{diekmann_inverse_2022} that low absorption in the cladding layer is favorable and potentially more important than high refractive index contrasts. This also includes cavities without upper cladding layer as an interesting alternative to traditional cavity designs.

For the direct inverse design of observable spectra, we employed the numerically efficient Green's function approach~\cite{lentrodt_ab_2020}, and introduced a representation of the reflection spectrum in a diagonal basis.  Using the latter tool, the spectrum can be characterized in terms of a few parameters that are readily accessible analytically and thus allow for an efficient numerical treatment. As an example, we again employed cavities with two resonant layers, which in general feature two Lorentzian resonances contributing to the spectrum. We then used the formalism to comprehensively determine the accessible frequency shifts and linewidths of these resonances. 

As a first result, we showed that we can design the two-layer system to exhibit spectra dominated by a single resonance, thereby mimicking the response of cavities with a single nuclear layer. However, the two-layer case has several advantages. First, the observable tuning range is larger, even for the same amount of resonant material in the two cavities. Second, cavities with a single layer of ${}^{57}$Fe are bound to low resonant layer thicknesses, as otherwise a magnetic long-range ordering appears, which gives rise to a magnetic splitting of the bare nuclear resonances. Third, the cavities with a single resonant layer feature the most extreme collective parameter modifications in cavities with vanishing visibility of the nuclear signatures in the observable spectra, which do not have an experimental relevance~\cite{diekmann_inverse_2022}. In contrast, the two-layer cavities can be designed to have  superior collective parameters which remain well-observable in the reflection spectrum.

Finally, we generalized the inverse design of the spectra, and showed that it allows one to tune the spectra towards arbitrary signatures within the physical limitations of the platform. To this end, we defined a cost function, which specifies the desired properties of the reflection spectrum. Maximizing this cost function then determines the optimal cavity structure to comply with the given design goals.
We illustrated this approach by designing cavities featuring two resonances with a tunable distance, relative height, or absolute position in the reflection spectrum. 

In the present work, we focused the examples to two resonant layers and an elementary surrounding thin-film cavity structure. However, our approach can readily be generalized to more complex scenarios, comprising more resonant or non-resonant layers. We envision that our approach of designing and optimizing artificial multi-level schemes will facilitate the implementation of a wider range of quantum optical schemes in the hard x-ray regime. Analogously, the inverse design of reflection spectra opens up new vistas for improved functionalities, and more layers give rise to an even higher adaptivity of the spectrum than in the examples reported here. However, with increasing system complexity, the realization of a desired functionality by trial and error approaches becomes harder, and a comprehensive methodology as provided here is all the more needed.

\begin{acknowledgments}
 OD gratefully acknowledges financial support by the Cusanuswerk, the Studienstiftung des Deutschen Volkes and the Austrian Science Fund (FWF) Grant No. P32300. DL gratefully acknowledges the Georg H. Endress Foundation for financial support.
\end{acknowledgments}

\appendix
\bibliography{reference.bib}
\bibliographystyle{myprsty}
\end{document}